\documentclass{IEEEtran}
\usepackage{epsfig}
\usepackage{amsmath}
\usepackage{amstext}
\usepackage{amsfonts}
\usepackage{amssymb}
\usepackage{eucal}
\usepackage{graphicx}
\usepackage{verbatim}
\usepackage{stmaryrd}
\usepackage{bm}

\usepackage{tikz}
\usepackage{pgfplots}
\usepackage{algorithmic}

\usetikzlibrary{arrows,petri,topaths}
\usepackage{tkz-berge}

\newcommand{\RR}{{\mathbb{R}}}

\newcommand{\CC}{{\mathbb{C}}}

\newcommand{\trans}{{\sf T}}

\newcommand{\PP}{{\mathbb{P}}}

\newcommand{\1}{{{1}}}
\newcommand{\interior}{{\rm{Int}}}

\newcommand{\oh}{{\frac{1}{2}}}

\newcommand{\asto}{\overset{\rm a.s.}{\longrightarrow}}
\newcommand{\probto}{\overset{\mathbb P}{\longrightarrow}}

\newcommand{\EE}{{\rm E}}

\DeclareMathOperator{\tr}{tr}
\DeclareMathOperator{\diag}{diag}

\newcounter{ctheorem}
\newtheorem{theorem}[ctheorem]{Theorem}

\newcounter{ccorollary}
\newtheorem{corollary}[ccorollary]{Corollary}
\newcounter{clemma}
\newtheorem{lemma}[clemma]{Lemma}

\newcounter{cremark}
\newtheorem{remark}[cremark]{Remark}

\begin{document}
\bibliographystyle{IEEEtran}

\title{Fluctuations of spiked random matrix models\\ and failure diagnosis in sensor networks}

\author{Romain~Couillet$^{1}$ and Walid~Hachem$^2$\\ {\it $^1$ Sup\'elec, Gif sur Yvette, France.} \\ {\it $^2$ CNRS-Telecom ParisTech, Paris, France.}}

\maketitle

\begin{abstract}
	In this article, the joint fluctuations of the extreme eigenvalues and eigenvectors of a large dimensional sample covariance matrix are analyzed when the associated population covariance matrix is a finite-rank perturbation of the identity matrix, corresponding to the so-called spiked model in random matrix theory. The asymptotic fluctuations, as the matrix size grows large, are shown to be intimately linked with matrices from the Gaussian unitary ensemble (GUE). When the spiked population eigenvalues have unit multiplicity, the fluctuations follow a central limit theorem. This result is used to develop an original framework for the detection and diagnosis of local failures in large sensor networks, for known or unknown failure magnitude. 
 \end{abstract}

 \section{Introduction}

 In the field of fault detection and diagnosis, one of the elementary requests is the fast, reliable and computationally light identification of a system failure. In dynamical scenarios, these systems are composed of several fluctuating parameters whose evolutions are tracked by a mesh of sensors reporting successive correlated and noisy data measurements to a central decision unit. With the growth in size and complexity of such systems, it becomes increasingly difficult for decision units to process simultaneously and at a low computational cost the augmenting load of reported measurements. Examples of such systems are the recent cognitive radio networks \cite{MIT99} and smart grid technologies \cite{AMI05}. In the former, multiple cooperative wireless communication devices, referred to as the secondary network, exchange sensed data in order to decide collectively which communication bandwidths are left unused by the licensed, also called primary, network users. Fast detection of sudden changes, e.g. new primary user communications, is here demanded to minimize the interference generated by secondary users. In the smart-grid framework, a large dimensional graph of interconnected electricity producers, transportation systems, and consumers evolve in real-time, their behaviour being reported by diverse sensors such as voltage phasor measurements \cite{PHA93} at the nodes of the electricity grid to regional controllers. Fast detection of link and node failures is requested in this scenario to minimize the risk of cascaded failures leading to regional blackouts \cite{KIN05}. There exists a rich literature on failure detection, diagnosis and change-point estimation, ranging from off-line detection methods of uncorrelated data \cite{BAS93,NIK95} to fast change detection methods in time correlated signals \cite{UNN11,VEE01,DAN09}. Subspace methods were in particular proposed to detect system changes from modifications in the eigenstructure of sampled covariance matrices for dynamical systems \cite{CHO05,BAS00}. In this article, we propose a novel subspace approach to solve the problem of off-line detection and identification of local failures from independent or linearly time-correlated samples. 

We precisely assume the observation of measurements suggesting an error has already occurred in the network. We wish the detection of a failure to be fast so we will assume that the number $n$ of successive sensor data reports is not extremely large with respect to the size $N$ of the network. We will also assume that the hypothetical failure scenarios are, to some extent, known in advance. In this context, calling $\mathcal H_0$ the hypothesis that the system does not undergo any failure and $\mathcal H_k$, $1\leq k\leq K$, the hypothesis that a failure of type $k$ occurs, the question of failure detection and localization consists in proceeding to the successive hypotheses tests: (i) decide whether the concatenation matrix of $n$ successive network observations $\Sigma=[s_1,\ldots,s_n]\in\CC^{N\times n}$ suggests hypothesis $\mathcal H_0$ or its complementary $\bar{\mathcal H}_0$ (i.e. the event union of the $\mathcal H_k$), and (ii) upon decision of $\bar{\mathcal H}_0$, decide what $\mathcal H_k$ is the most likely. Both problems are optimally solved by multi-hypothesis Neyman-Pearson tests \cite{NEY28} with maximum likelihood performance given the observation $\Sigma$. However, this procedure is computationally intense for large system dimensions and large $K$. 
 
The approach under consideration here follows the theory of large dimensional random matrices. Precisely, we consider the setting where both $N$ and $n$ grow large and such that $c_N=N/n\to c$, with $0<c<1$, as $N,n\to\infty$. Under this assumption, we develop asymptotic results on the extreme eigenvalues and associated eigenvectors of a certain family of random matrices to provide novel subspace methods for failure detection and localization. Our interest is on random matrices of the {\it spiked model} type, introduced by Johnstone \cite{JOH01}, specifically here of matrices modeled as $\Sigma=(I_N+P)^\oh X$, where $X$ is a left-unitarily invariant random matrix and $P$ is a rank-$r$ Hermitian matrix with $r\ll N$. Such matrix models have been largely studied in the recent random matrix literature, very often in the special case where $X$ is a {\it standard Gaussian matrix}, which refers in this article to a random matrix with independent $\mathcal{CN}(0,1/n)$ entries. In \cite{SIL98,BAI06}, for $X$ a standard Gaussian matrix, it is first shown that there exists a natural mapping between the extreme (empirical) eigenvalues of $\Sigma\Sigma^\ast$ and the (population) eigenvalues of $P$. It is then proved that, almost surely, the extreme empirical eigenvalues converge to deterministic limits in the asymptotic setting, found either at the edges of the support of the Mar\u{c}enko-Pastur law \cite{MAR67}, i.e. the (almost sure) weak limit of the eigenvalue distribution of $XX^\ast$, or away from them, depending on the corresponding population eigenvalues. This induces a phase transition having important consequences on fault detectability in sensor networks \cite{NAD10}. This observation is extended to the non-Gaussian case and generalized to other spiked models in \cite{BEN09}. The fluctuations of the extreme eigenvalues are studied with different approaches depending on whether the limiting eigenvalues are found at the edge or outside the support of the Mar\u{c}enko-Pastur law. When at the edge, it is proved successively in \cite{TRA96,JOH01,JOH00,BAI05,FEL10} that the (centered and scaled) limiting eigenvalue has Tracy-Widom fluctuations. When outside the support, those fluctuations are linked to the distribution of the eigenvalues of GUE matrices, as shown in \cite{BAI05}. In the specific case where the spiked eigenvalues of $P$ have unit multiplicity, the fluctuations are Gaussian.

In this article, the properties of the extreme eigenvalues in a spiked model will be used to provide failure detection tests, in the same line as \cite{CAR08,BIA10}. For failure localization, the information on the eigenvalue position can be used to reduce the number of hypotheses $K$. However, tests solely based on the limiting properties of the eigenvalues will turn out to be inefficient to discriminate the remaining hypotheses and we therefore develop novel results on the eigenspaces associated to these eigenvalues. In \cite{PAU07}, it is shown, in the real Gaussian case, that the projection of the eigenvectors associated with the extreme empirical eigenvalues of $\Sigma\Sigma^\ast$ on the subspace of the corresponding population eigenvectors of $P$ has a positive limiting norm, which is close to $1$ for $c$ small. This remark is extended in \cite{BEN09} to the non-Gaussian case. This property is the basis of our novel failure diagnosis method. However, the fluctuations of the eigenvector projections, fundamental here to derive test statistics for failure localization, have never been derived before for either the Gaussian or the non-Gaussian cases. The main mathematical result of this article, Theorem \ref{th:2ord}, provides the joint fluctuations of the eigenvalues and eigenspace projections for the eigenvalues found away from the limiting support of $XX^\ast$. Our proof technique is largely inspired by \cite{BEN09,ben-gui-mai-11,ben-rao-rectangular-11}. We also use some tools from \cite{pas-vas-book07} and \cite{hac-spike-11}. Based on these results, we suggest an original framework for local failure detection and identification in large sensor networks.

 The remainder of this document unfolds as follows. Section \ref{sec:failure} introduces elementary examples of sensor networks for which local failures translate into small rank perturbations of the identity matrix. Section \ref{sec:main_res} reminds important notions of random matrix theory and introduces the main mathematical results of this article. Practical application algorithms along with simulations are then carried out in Section \ref{sec:application}. Finally, Section \ref{sec:conclusion} concludes the article.

 {\it Notations:} In this document, capital characters stand for matrices while lowercase characters stand either for scalars or vectors, with $I_N\in\CC^{N\times N}$ the identity matrix. The $i^{\rm th}$ entry of a vector $x$ is denoted $x(i)$. The symbol $(\cdot)^\ast$ denotes complex transpose. For a function $f$ and a Hermitian matrix $X\in\CC^{N\times N}$, $f(X)=U\diag(f(\lambda_1(X)),\ldots,f(\lambda_N(X)))U^\ast$ with $\lambda_1(X),\ldots,\lambda_N(X)$ the eigenvalues of $X$ and $U\in\CC^{N\times N}$ the unitary matrix of its respective eigenvectors. The symbol $S_\pi$ denotes the support of the probability measure $\pi$. The notation ${\rm Span}(u_1,\ldots,u_k)$ designates the space generated by the vectors $u_1,\ldots,u_k$. The notation $S^\perp$ is the space orthogonal to $S$. We denote $\CC^+=\{z\in\CC,~\Im(z)>0\}$. The norm $\|X\|$ of a Hermitian matrix $X$ is understood as the spectral norm, and the norm $\|x\|$ of a vector $x$ is understood as the Euclidean norm. The notations `$\asto$', `$\Rightarrow$', and `$\probto$' denote convergence almost surely, weakly, and in probability, respectively. The symbol $\EE[\cdot]$ denotes expectation. The notation $1_A(x)$ is the indicator function on the set $A$.

 \section{Detection and localization of local failures}
 \label{sec:failure}

 To motivate the study of the fluctuations of extreme eigenvalues and eigenvectors of sample covariance matrices in the context of local failures in large dimensional sensor networks, we introduce in the following two basic examples of sensor network failure scenarios, which can all be modeled as small rank perturbations of the identity matrix, as well as related engineering applications.

 \subsection{Node failure}
 \label{sec:ex1}
 Consider the following model 
 \begin{equation}
	 \label{eq:model1}
	 y = H \theta + \sigma w
 \end{equation}
 where $H\in\CC^{N\times p}$ is deterministic, $\theta=[\theta(1),\ldots,\theta(p)]^\trans\in\CC^p$, $w\in\CC^N$ have independent and identically distributed (i.i.d.) complex standard Gaussian entries, and $\sigma>0$. We denote $y=[y(1),\ldots,y(N)]^\trans\in\CC^N$. In a sensor network composed of $N$ nodes, $y$ represents the observation through the channel $H$ of the vector $\theta$, constituted of centered and normalized independent Gaussian system parameters,\footnote{Up to a right-product of $H$ by a positive diagonal matrix, the variance of the entries of $\theta$ can be assumed all equal to one without loss of generality.} impaired by white Gaussian noise. Therefore, $\EE[yy^\ast]=HH^\ast + \sigma^2I_N\triangleq R$. 
 
 In case of failure of sensor $k$, $y(k)$, the $k^{th}$ entry of $y$, will start suddenly to return noisy outputs inconsistent with the model \eqref{eq:model1}. Assuming this noise Gaussian with zero mean and variance $\sigma_k^2$ and denoting $y'$ the observations of the network with failure at sensor $k$, we can write
 \begin{equation*}
y' = (I_N-e_ke_k^\ast)H\theta + \sigma_k e_k e_k^\ast \theta' + \sigma w
 \end{equation*}
where $\theta'$ is distributed like $\theta$ but independently and $e_k\in\CC^N$ is such that $e_k(k)=1$ and $e_k(i)=0$, for all $i\neq k$.

Therefore, $y'$ is Gaussian (as the sum of Gaussian variables) with zero mean and variance
\begin{equation*}
	\EE[y'y^{\prime\ast}] = (I_N-e_ke_k^\ast)HH^\ast(I_N-e_ke_k^\ast) + \sigma_k^2 e_k e_k^\ast + \sigma^2 I_N.
\end{equation*}
Denoting $s=R^{-\oh}y'$, we have
\begin{align*}
	\EE[ss^{\ast}] &= I_N  - R^{-\oh} HH^\ast e_k e_k^\ast R^{-\oh} \\
	& + R^{-\oh}e_k \left[(e_k^\ast HH^\ast e_k + \sigma_k^2)e_k^\ast R^{-\oh} - e_k^\ast HH^\ast R^{-\oh}\right].
\end{align*}
Therefore, the population covariance matrix $\EE[ss^{\ast}]$ is a perturbation of the identity matrix by
\begin{align}
	\label{eq:Pk}
	P_{k} &\triangleq R^{-\oh}e_k \left[(e_k^\ast HH^\ast e_k + \sigma_k^2)e_k^\ast R^{-\oh} - e_k^\ast HH^\ast R^{-\oh}\right] \nonumber \\ 
	&- R^{-\oh} HH^\ast e_k e_k^\ast R^{-\oh}.
\end{align}
Notice that the image of $P_{k}$ is included in the subspace ${\rm Span}(R^{-\oh}e_k,R^{-\oh} HH^\ast e_k)$ and is therefore at most of dimension two.
Generalizing the above to $M$ node failures at nodes $k_1,\ldots,k_M$, the vector $s$ is now such that
\begin{align*}
	\EE[ss^{\ast}] &= I_N - R^{-\oh} HH^\ast EE^\ast R^{-\oh} \\ 
	&+ R^{-\oh}E \left[ (E^\ast HH^\ast E + \Lambda^2)E^\ast R^{-\oh} - E^\ast HH^\ast R^{-\oh}\right] 
\end{align*} 
with $E=[e_{k_1},\ldots,e_{k_M}]$, $\Lambda = \diag(\sigma_{k_1},\ldots,\sigma_{k_M})$, where now \eqref{eq:Pk} becomes
\begin{align}
	\label{eq:Pk1kr}
	P_{k_1,\ldots,k_M} &\triangleq R^{-\oh}E \left[ (E^\ast HH^\ast E + \Lambda^2)E^\ast - E^\ast HH^\ast\right]R^{-\oh} \nonumber \\
	&- R^{-\oh}HH^\ast EE^\ast R^{-\oh}
\end{align} 
which is a matrix of rank at most $2M$.

\subsection{Sudden parameter change}
\label{sec:ex2}
Consider again the elementary model \eqref{eq:model1} and now assume that, instead of a sensor failing, $\theta(k)$, the $k^{th}$ entry of $\theta$, experiences a sudden change in mean and variance. The resulting observation $y'$ can be modeled as
\begin{equation*}
y' = H(I_p+\alpha_k e_ke_k^\ast)\theta + \mu_k He_k + \sigma w 
\end{equation*}
for some real parameters $\mu_k,\alpha_k$, and where $e_k\in\CC^p$ is defined by $e_k(k)=1$ and $e_k(i)=0$, $i\neq k$. For this particular model, we may or may not suppose that $\mu_k$ and $\alpha_k$ are a priori known to the experimenter. In this scenario, we now have that $y'$ is Gaussian with zero mean and variance
\begin{equation*}
	\EE[y'y^{\prime\ast}] = H(I_p+[\mu_k^2+(1+\alpha_k)^2-1] e_ke_k^\ast)H^\ast + \sigma^2 I_N. 
\end{equation*}
Denoting $R=HH^\ast+\sigma^2I_N$ as in the previous example and taking $s=R^{-\oh}y'$, we finally have
\begin{equation*}
	\EE[ss^\ast] = I_N + [\mu_k^2+(1+\alpha_k)^2-1] R^{-\oh}He_ke_k^\ast H^\ast R^{-\oh}
\end{equation*}
which is a rank-$1$ perturbation of the identity matrix by the matrix
\begin{equation*}
	P_k \triangleq \beta_k R^{-\oh}He_ke_k^\ast H^\ast R^{-\oh}
\end{equation*}
with $\beta_k=\mu_k^2+(1+\alpha_k)^2-1$. Note that, in this scenario, the eigenvector of $P_k$ associated with the non-zero eigenvalue is independent of $\mu_k$ and $\alpha_k$. For practical applications, this has the interesting advantage that simple localization can be performed even if $\mu_k$ and $\alpha_k$ are unknown. This is further discussed in Section \ref{sec:application}.

The derivation above generalizes to sudden changes of multiple parameters. If the means and variances for the sensors $k_1,\ldots,k_M$ are modified simultaneously with respective parameters $\mu_{k_1},\ldots,\mu_{k_M}$ and $\alpha_{k_1},\ldots,\alpha_{k_M}$, then
\begin{equation*}
	\EE[ss^\ast] = I_N + R^{-\oh}HE\Lambda E^\ast H^\ast R^{-\oh}
\end{equation*}
with $E=[e_{k_1},\ldots,e_{k_M}]$ and $\Lambda=\diag(\beta_{k_1},\ldots,\beta_{k_M})$, $\beta_{k_i}=\mu_{k_i}^2+(1+\alpha_{k_i})^2-1$, which is a rank-$M$ perturbation of the identity matrix by the matrix
\begin{equation*}
	P_{k_1,\ldots,k_M} = R^{-\oh}HE\Lambda E^\ast H^\ast R^{-\oh}. 
\end{equation*}
Note that, contrary to the one-dimensional case, the eigenvectors of $P_{k_1,\ldots,k_M}$ depend here explicitly on the parameters $\mu_{k_i}$ and $\alpha_{k_i}$.

In the following section, we introduce the novel detection and localization framework and we discuss engineering applications referencing the examples described in this section.

\subsection{Detection and localization in sensor networks}

For either of the models above, let us assume a general scenario with $K$ possible failure events, indexed by $1\leq k\leq K$ and let now $s_1,\ldots,s_n$ be $n$ successive independent observations of the random variable $s$. We denote $\Sigma\triangleq \frac1{\sqrt{n}}[s_1,\ldots,s_n]\in\CC^{N\times n}$. From the fact that $s$ is Gaussian with zero mean and covariance $(I_N+P_k)$ for a certain $k$, we can write
\begin{equation*}
	\Sigma=(I_N+P_k)^\oh X
\end{equation*}
where $X\in\CC^{N\times n}$ is a given matrix with independent Gaussian entries of zero mean and variance $1/n$. We also denote for simplicity $P_0=0$ for the extra event $k=0$ corresponding to the no-failure scenario. 

The natural approach to detect and identify a failure event in a sensor network upon the observations $s_1,\ldots,s_n$ is to systematically perform a maximum likelihood test on the $K+1$ hypotheses $\mathcal H_0,\ldots,\mathcal H_K$, with $\mathcal H_k$ defined as the event $s \sim \mathcal{CN}(0,I_N+P_k)$.
However, this optimal approach has some intrinsic limitations. From a computational aspect, evaluating the probability of each hypothesis $k$ requires to evaluate the term $\tr \Sigma^\ast(I_N+P_k)^{-1}\Sigma$, an operation whose cost is of order $\mathcal O(N^3)$ (which can be brought down to $\mathcal O(N^2)$ using matrix inversion lemmas). When the number of hypotheses $K$ and the system size $N$ are large, these operations become extremely demanding. Pre-calculus of the inverses $(I_N+P_k)^{-1}$ also requires possibly large memory storage.

Since the node failure information is entirely captured by the perturbation matrix $P_k$, we provide in the following a suboptimal test relying on the properties linking $P_k$ to the observation matrix $\Sigma$, for large system dimensions $(N,n)$. Precisely, based on recent advances in the field of large dimensional random matrix theory \cite{COUbook}, we provide a two-step approach to successively (i) decide on the existence of a failure from the location of the extreme eigenvalues of $\Sigma\Sigma^\ast$ and (ii) identify the failure event from eigenspace projections. This diagnosis framework relies on the asymptotic statistics of these extreme eigenvalues and eigenspace projections. This subspace approach has multiple advantages compared to the optimal hypothesis testing method discussed above. From a computational aspect, step (i) requires to determine the eigenvalues of $\Sigma\Sigma^\ast$, hence a singular value decomposition. This step already provides sufficient information for step (ii) to become computationally cheap: on the one hand, the position of the extreme eigenvalues of $\Sigma\Sigma^\ast$ may be used to reduce the set $\mathcal H_1,\ldots,\mathcal H_K$ to a possibly small subset of consistent hypotheses; on the other hand, for the remaining hypotheses, the localization test will merely consists in the characterization of eigenvector projections, an operation of computational cost $\mathcal O(N)$. No matrix inverse needs to be computed and only the eigenvectors and non-zero eigenvalues of $P_k$ need to be stored. The technique also has the advantage to be consistent in its usage of eigenvalues and eigenspace projections to perform hypothesis tests. Finally, as will be discussed in Section \ref{sec:application}, the framework can be extended to account easily for unknown failure amplitudes, which would be much more involved from a maximum-likelihood approach.

Before presenting our main results, we briefly discuss practical applications in the failure diagnosis context.
The application of the aforementioned results to failure diagnosis in sensor networks consists in deriving subspace methods for failure detection and identification when the system undergoes sudden changes that can be modeled as small rank perturbations of the data covariance matrix. The scenario of Section \ref{sec:ex1} may be exploited in particular to rapidly detect failures in sensors that may suddenly return inconsistent data. Fast detections of such sensor errors are of fundamental importance when decisions on system actuators are taken from these data. The scenario of Section \ref{sec:ex2} finds applications in the cognitive radio setting where Gaussian baseband signals $\theta(1),\ldots,\theta(p)$ arising from a set of $p$ fixed access points, forming the so-called primary network, are sensed by a network composed of $N$ cooperating sensors, constituting the secondary network. Assuming that at a reference time instant a subset $\mathcal P$ of the $p$ access points transmit data, the communication channel $H\in\CC^{N\times p}$ between primary and secondary networks (which can be assumed static during a channel coherence interval) will contain non-zero entries in the columns indexed by $\mathcal P$. The objective of the secondary network is here to rapidly detect and identify changes in $H$ (column elimination or creation), corresponding to evolutions of the subset $\mathcal P$ of transmitting access points. These fast change diagnoses will be used to exploit the available radio spectrum as well as to avoid secondary transmissions to interfere primary communications. We also mention that in \cite{COUZIO}, a similar system model for failure diagnosis in power systems is analyzed. In this setting, the observation vector $y$ is composed of voltage measurements at the $N$ network nodes, the variable $\theta$ contains the current inputs at the nodes which naturally vary due to the presence of unreliable renewable energy production units, and the transfer matrix $H$ is closely related to the square matrix containing at entry $(i,j)$ the inverse impedance of the power line connecting $i$ to $j$ (and equal to zero if $i$ and $j$ are not connected). The objective in this system is to detect local power line outages corresponding to impedance changes in $H$, which can be seen as rank-two perturbations of the population covariance matrix.

The following section is dedicated to the study of the asymptotic eigenvalue and eigenspace projection statistics as the dimensions of the matrix $\Sigma$ grow large. 

\section{Main results}
\label{sec:main_res}
The derivation arguments found in this section follow the ideas of \cite{ben-rao-advmath-11}, \cite{ben-gui-mai-11}, and \cite{ben-rao-rectangular-11}. In our proofs, we shall also borrow some of the arguments of \cite{hac-spike-11} whose context is close to ours. 

 \subsection{Notations, assumptions and basic results}
 \label{sec:notations}

We start by summarizing the major notations and facts needed here. We consider a generic small rank perturbation model and define
 \begin{align*}
	 \Sigma = (I_N+P)^\oh X
 \end{align*}
with $X\in \CC^{N\times n}$ left-unitarily invariant, and where the rank-$r$ Hermitian matrix $P$ has the spectral factorization $P=U\Omega U^\ast$ with 
\begin{equation*}
\Omega = \begin{bmatrix} \omega_1 I_{j_1} &  \\
& \ddots \\
& & \omega_t I_{j_t} \end{bmatrix}
\end{equation*}
and $\omega_1 > \ldots > \omega_s > 0 > \omega_{s+1} > \ldots > \omega_t > -1$. Of course, $j_1+\cdots+j_t = r$. We write accordingly $U = \begin{bmatrix} U_1 \cdots U_t \end{bmatrix}$ where $U_i\in\CC^{N\times j_i}$. We denote by $\hat\lambda_1 \geq \cdots \geq \hat\lambda_N$ the eigenvalues of $\Sigma\Sigma^*$. For $i \in \{ 1, \ldots, s \}$, we let ${\mathcal K}(i) = j_1 + \cdots + j_{i-1}$, taking by convention $j_0 = 0$. For $i \in \{ s+1,\ldots, t+1\}$, we let ${\mathcal K}(i) = N - (j_i+\cdots+j_t)$. 
One of the purposes of this section is to establish an asymptotic relation between $\omega_i$ and the $\hat\lambda_{{\mathcal K}(i)+\ell}$ for $\ell=1,\ldots, j_i$ which holds under a condition on $\omega_i$ that will be specified. We also denote by $\widehat\Pi_{i}$ the orthogonal projection matrix, when it exists, on the eigenspace of $\Sigma\Sigma^*$ associated with the eigenvalues $\{ \hat\lambda_{{\mathcal K}(i)+\ell} \}_{\ell=1}^{j_i}$. Similarly, we denote by $\Pi_i = U_i U_i^*$ the orthogonal projection matrix on the eigenspace of $P$ associated with the eigenvalue $\omega_i$. Finally, we denote by $Q(z) = (XX^\ast-z I_N)^{-1}$ the resolvent of the matrix $XX^\ast$ and by $\alpha(z) = \frac1N \tr Q(z)$ its normalized trace, both analytical on $\CC^+$.

In the remainder of the paper, we shall consider the asymptotic regime where $n\to\infty$ and $N/n \to c \in (0,1)$. The notation $n\to\infty$ will henceforth refer to this asymptotic 
regime.

We now state our basic assumptions:
\begin{description} 
\item[{\bf A1}] 
The probability law of $X$ is invariant by left multiplication by a deterministic unitary matrix. 
\end{description}
Thanks to the left unitary invariance of $X$, $Q(z)$ writes as $Q(z) = W (\Lambda - z I_n)^{-1} W^*$ where $\Lambda$ is the matrix of eigenvalues of $XX^\ast$, $W$ is a unitary random matrix Haar distributed on its unitary group, and $W$ and $\Lambda$ are independent. 
\begin{description} 
\item[{\bf A2}] 
For every $z \in \CC^+$, $\alpha(z)$ converges almost surely to a deterministic function $m(z)$ which is the Stieltjes transform\footnote{We recall that the Stieltjes transform $m(z)$ of a real measure $\pi$ is defined for $z$ outside the support of $\pi$ by $m(z)=\int \frac1{\lambda-z}d\pi(\lambda)$.} of a probability measure $\pi$ with support $[a,b] \subset (0,\infty)$.
\item[{\bf A3}] 
We have $\| X X^\ast \| \asto b$ and $(\| (XX^\ast)^{-1} \|)^{-1} \asto a$. 
\end{description}
This last assumption implies in particular that {\bf A2} is satisfied for all $z \in \CC \setminus [a,b]$. 

The most classical model of a matrix $X$ that satisfies {\bf A1}-{\bf A3} is when $X$ is standard Gaussian, i.e. with independent ${\mathcal CN}(0,1/n)$ elements, as introduced in the system models of Section \ref{sec:ex1} and Section \ref{sec:ex2}. For this model, the limiting probability distribution $\pi$ is the well known Mar\u{c}enko-Pastur distribution \cite{MAR67}. Its Stieltjes transform $m(z)$ is given by
\begin{equation}
\label{eq:mp} 
m(z) = \frac1{2zc} \left(1-c - z + \sqrt{(1-c-z)^2-4zc} \right)
\end{equation} 
where the branch of the square root is chosen such that $m(\CC^+)\subset \CC^+$ and $m$ is analytic on $\CC\setminus [a,b]$, where $a=(1-\sqrt{c})^2$ and $b=(1+\sqrt{c})^2$. 

The unitary invariance of $X$ is the basis of the following important lemma, shown in \cite{hac-spike-11} using an inequality of \cite{pas-vas-book07} which involves Haar unitary matrices: 
\begin{lemma}
\label{lm-fq} 
Assume {\bf A1}. Let $u\in\CC^N$ and $v\in\CC^N$ be two vectors with norm $\| u \| = \| v \| = 1$. Denote by $\sigma(XX^\ast)$ the eigenvalue spectrum of $XX^\ast$. Given $\varepsilon > 0$ and $z \in \CC \setminus [a-\varepsilon,  b+\varepsilon]$, denote by $d_z$ the distance from $z$ to $[a-\varepsilon, b+\varepsilon]$. Then for any $p > 0$, 
\begin{equation*}
\EE \left[\left| \1_{\sigma(XX^\ast) \subset [a-\varepsilon, b+\varepsilon]} 
u^\ast ( Q(z) - \alpha(z) I_N ) v \right|^p\right] 
\leq \frac{K_p}{d_z^p N^{p/2}} 
\end{equation*}
where the constant $K_p$ depends on $p$ only. Similarly, for any $z, z' \in \CC \setminus [a-\varepsilon,  b+\varepsilon]$, we have
\begin{align*} 
	&\EE \left[\left| \1_{\sigma(XX^\ast) \subset [a-\varepsilon, b+\varepsilon]} u^\ast \left[ Q(z)Q(z') - \frac{\tr Q(z) Q(z')}N I_N \right] v \right|^p\right] \\ 
&\leq \frac{K_p}{d_z^p d_{z'}^p N^{p/2}} . 
\end{align*} 
\end{lemma} 

We now start our analysis of the extreme eigenvalues and eigenspace projections of $\Sigma\Sigma^\ast$ by studying the first order behavior.

\subsection{First order behavior}
\label{sec:1storder} 

\subsubsection{Eigenvalues} 
\label{sec:vap}

Suppose that $x \in \RR$ is not an eigenvalue of $XX^*$. We first write 
\begin{align*}
&\det(\Sigma \Sigma^\ast - x I_N) \\
&= \det(I_N+P)\det(XX^\ast - xI_N + x[I_N-(I_N+P)^{-1}])\\
&= \det(I_N+P)\det(XX^\ast - xI_N) \\
&\times\det(I_N + x P(I_N+P)^{-1}(XX^\ast - xI_N)^{-1} )
\end{align*}
after noticing that $I_N-(I_N+P)^{-1}=P(I_N+P)^{-1}$. Therefore, if $x$ is an eigenvalue of $\Sigma\Sigma^\ast$ but not of $XX^\ast$, it must cancel the rightmost determinant.
This determinant can be further rewritten
\begin{align*} 
&\det(I_N + x P(I_N+P)^{-1}Q(x)) \\
&= \det (I_r + x\Omega U^\ast(I_N+U\Omega U^\ast)^{-1} Q(x)U).
\end{align*} 
From the identity $U^\ast(I_N+U\Omega U^\ast)^{-1}=(I_r+\Omega)^{-1}U^\ast$, we then have
\begin{equation*}
\det(\Sigma \Sigma^\ast - x I_N) = 
\det(I_N+P)\det(XX^\ast - xI_N)\det(\widehat{H}(x)) 
\end{equation*}
where 
$\widehat{H}(z)=I_r + z\Omega(I_r+\Omega)^{-1}U^\ast Q(z) U$. \\ 
Given any $z \in \CC \setminus [a,b]$, Assumptions {\bf A1}-{\bf A3} in conjunction with Lemma \ref{lm-fq} show that $\widehat{H}(z)$ is defined for all $n$ large, almost surely, and converges with probability one to 
\begin{equation}
	\label{eq:Hz}
H(z) = I_r + z m(z) \Omega(I_r+\Omega)^{-1}  
\end{equation}
(take $u$ and $v$ in Lemma \ref{lm-fq} as any couple of columns of $U$, take $p>2$ and use Borel-Cantelli's lemma \cite{BIL08}). We therefore expect the solutions of the equation $\det H(x) = 0$ which are outside $[a,b]$ to coincide with the limits of the isolated eigenvalues of $\Sigma\Sigma^\ast$.

Let us now study the behavior of the solutions of this equation. Let $h(x) = x m(x)$ on $\RR \setminus [a,b]$. Since $a>0$, we have
\begin{equation*}
h'(x) = \left(xm(x)\right)'=\int \frac{\lambda}{(\lambda-x)^2}
d\pi(\lambda) >0 . 
\end{equation*}
The function $h(x)$ is therefore increasing on $\RR \setminus [a,b]$ and with limit $0$ as $x\to 0$ and $-1$ as $x\to\infty$. Therefore, for $\omega_i>0$, \eqref{eq:Hz} leads to  
\begin{equation}
\label{eq:rhok}
h(\rho) + \frac{1+\omega_i}{\omega_i} = 0
\end{equation}
having a unique real solution $\rho_i$ satisfying $\rho_i>b$ if and only if $h(b^+) +({1+\omega_i})/{\omega_i}<0$.\footnote{We denote by $x^+$ and $x^-$ any quantity infinitesimally greater and smaller than the real $x$, respectively.} When $\omega_i<0$, \eqref{eq:rhok} has a unique solution $0<\rho_i<a$ if and only if $h(a^-)+({1+\omega_i})/{\omega_i}>0$. We therefore have the following result, for which a rigorous proof is found in \cite{ben-rao-advmath-11}: 
\begin{theorem}
\label{th-spk} 
Assume {\bf A1}-{\bf A3}. Let $p$ be zero or the maximum index such that $\omega_p > 0$ and $h(b^+)+({1+\omega_p})/{\omega_p}<0$. For $i=1,\ldots,p$, let $\rho_i$ be the unique solution of \eqref{eq:rhok} such that $\rho_i > b$. Then, 
\begin{equation*}
	\hat\lambda_{{\mathcal K}(i)+\ell} \asto \rho_{i}
\end{equation*}
for $i=1,\ldots,p$ and $\ell = 1,\ldots, j_i$, while 
\begin{equation*}
	\hat\lambda_{{\mathcal K}(p+1)+1} \asto b.
\end{equation*}
Let $q$ be $t+1$ or the minimum index such that $\omega_q < 0$ and $h(a^-)+({1+\omega_q})/{\omega_q}>0$. For $i=q,\ldots,t$, let $\rho_i$ be the unique solution of \eqref{eq:rhok} such that $\rho_i < a$. Then,
\begin{equation*}
	\hat\lambda_{{\mathcal K}(i)+\ell} \asto \rho_{i}
\end{equation*}
for $i=q,\ldots,t$ and $\ell = 1,\ldots, j_i$, while 
\begin{equation*}
	\hat\lambda_{{\mathcal K}(q)} \asto a.
\end{equation*}
\end{theorem}

In the remainder of the article, the variables $\omega_1,\ldots,\omega_p$ and $\omega_q,\ldots,\omega_t$ satisfying the conditions of Theorem \ref{th-spk} will be said to {\it satisfy the separation condition}. 

When $X$ is standard Gaussian, applying Theorem \ref{th-spk} shows after some simple derivations the following result: 
\begin{corollary}
	\label{co-spk}
	Consider the setting of Theorem \ref{th-spk}. Assume additionally that $X$ is standard Gaussian. Let $p$ be zero or the maximum index for which $\omega_p > \sqrt{c}$ and $q$ be $t+1$ or the minimum index such that $\omega_q < -\sqrt{c}$. Then 
\begin{equation} 
	\label{eq:rhoMP}
	\hat\lambda_{ {\mathcal K}(i)+\ell} \asto \rho_i = 1+\omega_{i} + c\frac{1+\omega_{i}}{\omega_{i}}
\end{equation}
for $i \in \{ 1,\ldots p, q, \ldots, t\}$, $\ell = 1,\ldots, j_i$, while 
\begin{align*}
	\hat\lambda_{{\mathcal K}(p+1)+1} &\asto (1+\sqrt{c})^2 \\
	\hat\lambda_{{\mathcal K}(q)} &\asto (1-\sqrt{c})^2.
\end{align*}
\end{corollary}

Corollary \ref{co-spk} implies that, for $\omega_i$ sufficiently far from zero (either positive or negative) or, equivalently, for $c$ sufficiently small, the spectrum of $\Sigma\Sigma^\ast$ exhibits $j_i$ eigenvalues outside the support $S_\pi$ of the Mar\u{c}enko-Pastur law $\pi$ which all converge to $\rho_{i}$. For failure detection purposes, upon observation of $\Sigma$, we may then test the null hypothesis $\Sigma=X$ (call it hypothesis $\mathcal H_0$) against the hypothesis $\Sigma=(I_N+P)^\oh X$ (call it hypothesis $\bar{\mathcal H}_0$), depending on whether eigenvalues of $\Sigma\Sigma^\ast$ are found outside $S_\pi$. Depending on the scenario, for $c$ small enough, it may be that a mere evaluation of the number of eigenvalues outside the support suggests the number of simultaneous failures in the sensor network. This is the case of the two failure scenarios described in Section \ref{sec:ex1} and Section \ref{sec:ex2}. However, the information on the extreme eigenvalues of $\Sigma\Sigma^\ast$, if sufficient for failure detection purposes, is usually not good enough to perform accurate failure localization. This is because different failure scenarios, characterized by different perturbation matrices $P$, may exhibit very similar eigenvalues. Also, if the failure amplitude is a priori unknown, then eigenvalues are in general irrelevant to discriminate between failure hypotheses; see the application Section \ref{sec:unknown_failure}. In such scenarios, we then need to consider eigenspace properties of $P$. This is the target of the following section.

\subsubsection{Projections on eigenspaces}
\label{sec:vep}

Given $i\leq t$, we now assume that $\omega_i$ satisfies the separation condition. Given two deterministic vectors $b_1,b_2\in\CC^N$ with bounded Euclidean norms, our purpose is to study the asymptotic behavior of $b_1^* \widehat \Pi_i b_2$. We shall show that this bilinear form is simply related with $b_1^* \Pi_i b_2$ in the asymptotic regime.

Our starting point is to express $b_1^* \widehat \Pi_i b_2$ as a Cauchy integral \cite{RUD86}. Denoting $\mathcal C_i$ a positively oriented contour encompassing only the eigenvalues $\hat\lambda_{{\mathcal K}(i) + \ell}$ of $\Sigma\Sigma^*$ for $\ell=1,\ldots, j_i$, we have
\begin{align*}
	&b_1^\ast \widehat \Pi_i b_2 \nonumber \\
	&= -\frac{1}{2\pi \imath} 
  \oint_{\mathcal C_i} 
   b_1^\ast (\Sigma\Sigma^\ast-zI_N)^{-1} b_2 \ dz \\
	&= -\frac{1}{2\pi \imath} \oint_{\mathcal C_i} 
     b_1^\ast(I_N+P)^{-\oh} \nonumber \\ 
	&\times
   \left[XX^\ast-zI_N+zP(I_N+P)^{-1}\right]^{-1}(I_N+P)^{-\oh}b_2 \ dz.
\end{align*}
Using Woodbury's matrix identity, we have
\begin{align*}
	&\left[XX^\ast-zI_N+zP(I_N+P)^{-1}\right]^{-1} \\
	&=Q(z) - zQ(z)U\left[I_r+z\Omega(I_r+\Omega)^{-1}U^\ast Q(z) U\right]^{-1} \nonumber \\ 
	&\times \Omega(I_r+\Omega)^{-1} U^\ast Q(z)  \\ 
	&\triangleq Q(z) - zQ(z)U  \widehat H(z)^{-1} \Omega(I_r+\Omega)^{-1} U^\ast Q(z) 
\end{align*}
and taking
\begin{align*}
	\hat{a}_1(z)^\ast &= z b_1^\ast(I_N+P)^{-\oh} Q(z)U \\
	\hat{a}_2(z) &= \Omega(I_r+\Omega)^{-1} U^\ast Q(z)(I_N+P)^{-\oh}b_2
\end{align*}
we obtain
\begin{align}
	&b_1^\ast \widehat \Pi_i b_2 \nonumber \\
	&= -\frac{1}{2\pi \imath} 
\oint_{\mathcal C_i}
   b_1^\ast (I_N+P)^{-\oh} Q(z)(I_N+P)^{-\oh}b_2 \ dz \nonumber \\ 
	&+ \frac{1}{2\pi \imath} \oint_{\mathcal C_i}
     \hat{a}_1(z)^\ast \widehat{H}(z)^{-1}\hat{a}_2(z) \ dz.
\label{bpib} 
\end{align}
By Assumption {\bf A3} and Theorem \ref{th-spk}, with probability one for all large $n$, the first term on the right-hand side is zero, while the second is equal to  
\begin{align*}
\frac{1}{2\pi \imath} \oint_{\gamma_i}
\hat{a}_1(z)^\ast \widehat{H}(z)^{-1}\hat{a}_2(z) dz
\end{align*}
where $\gamma_i$ is a deterministic positively oriented circle away from $[a,b]$ enclosing $\rho_i$ but none of the $\rho_j$, $j\neq i$. Using Lemma \ref{lm-fq} in conjunction with the analyticity properties of the integrand, one can show that $\hat{a}_1(z)^\ast \widehat{H}(z)^{-1}\hat{a}_2(z)$ converges
uniformly to $a_1(z)^\ast H(z)^{-1}a_2(z)$ on $\gamma_i$ in the almost sure sense, where 
\begin{align*}
	a_1(z)^\ast &= zm(z) b_1^\ast(I_N+P)^{-\oh}U \\
	a_2(z) &= m(z) \Omega(I_r+\Omega)^{-1} U^\ast (I_N+P)^{-\oh}b_2 . 
\end{align*}
It results that $b_1^\ast \widehat \Pi_i b_2 - T_i \asto 0$, where 
\begin{align*}
T_i \triangleq \frac{1}{2\pi \imath} \oint_{\gamma_i} 
  a_1(z)^\ast H(z)^{-1}a_2(z) dz.
\end{align*}
Details can be found in \cite{hac-spike-11} in a similar situation. Let us find the expression of $T_i$. Noticing that
\begin{equation}
\label{eq:H-1} 
H(z)^{-1} = \sum_{\ell=1}^t 
\frac1{1+zm(z)\frac{\omega_\ell}{1+\omega_\ell}} {\cal I}_\ell
\end{equation}
where 
\begin{equation*}
	{\cal I}_\ell = \begin{bmatrix} 
		0_{j_1+\ldots+j_{\ell -1}} & \\
& I_{j_\ell} \\ 
& & 0_{j_{\ell +1}+\ldots+j_t} 
\end{bmatrix} \in \CC^{r\times r}
\end{equation*} 
we obtain 
\begin{align*}
	T_i &= \sum_{\ell=1}^t 
\frac{\omega_\ell}{(1+\omega_\ell)^2} 
b_1^\ast \Pi_\ell b_2 
\frac1{2\pi \imath}\oint_{\gamma_i} 
\frac{zm^2(z)}{1+zm(z)\frac{\omega_\ell}{1+\omega_\ell}}dz \\ 
&= \sum_{\ell=1}^t 
\frac{b_1^\ast \Pi_\ell b_2}{1+\omega_\ell} 
\frac1{2\pi \imath}\oint_{\gamma_i} 
\frac{zm^2(z)}{\frac{1+\omega_\ell}{\omega_\ell}+zm(z)}dz . 
\end{align*}
From the discussion prior to Theorem \ref{th-spk}, it is clear that the denominator of the integrand has a zero in the interior $\interior(\gamma_i)$ of the disk delineated by $\gamma_i$ only when $\ell=i$, and then this zero is simple due to the fact that $h'(x) = (xm(x))'$ never vanishes on $\RR\setminus[a,b]$. Applying the residue theorem \cite{RUD86} and observing from \eqref{eq:rhok} that $1/(1+\omega_i) = (1+h(\rho_i))/h(\rho_i)$, we obtain the following limits:

\begin{theorem}
\label{th-bf}
Assume {\bf A1}-{\bf A3}. Given $i \leq t$, assume that $\omega_i$ satisfies the separation condition. Let $b_1\in\CC^N$ and $b_2\in\CC^N$ be two sequences of increasing size deterministic vectors with bounded Euclidean norms. Then 
\begin{equation*}
b_1^* \widehat \Pi_i b_2 - \zeta_i b_1^\ast \Pi_i b_2 \asto 0 
\end{equation*}
where 
\begin{equation*}
\zeta_i = \frac{m(\rho_i)(1+h(\rho_i))}{h'(\rho_i)}. 
\end{equation*}
\end{theorem} 

In particular, we find after some derivations: 
\begin{corollary}
	\label{co-bf}
	Under the assumptions of Theorem \ref{th-bf}, let $X$ be standard Gaussian. Then 
\begin{equation*}
b_1^* \widehat \Pi_i b_2 - \frac{1-c\omega_i^{-2}}{1+c\omega_i^{-1}} b_1^\ast \Pi_i b_2 \asto 0. 
\end{equation*}
\end{corollary}
This result is consistent with \cite{PAU07} derived in the real Gaussian case for eigenvalues with unit multiplicity.

Theorem \ref{th:2ord} and Corollary \ref{co-bf2} provide an interesting characterization of the eigenspaces of $P$ through limiting projections in the large dimensional setting. In the context of local failure in large sensor networks, it is therefore possible to detect and diagnose one or multiple failures by comparing eigenspace projection patterns associated with each failure type. Precisely, an appropriate diagnosis consists in determining the most likely failure type among all hypothetical failures, given the extreme eigenvalues and associated eigenspace projections of $\Sigma\Sigma^\ast$. To this end though, not only first order limits but also second order behaviour need be characterized precisely. This is the target of the following section. 

\subsection{Second order behavior}
\label{sec:2ord}
Before studying the fluctuations of $\hat{\lambda}_{\mathcal K(i)+\ell}$, $\ell=1,\ldots,j_i$, when $\omega_i$ satisfies the separation condition, we first remind for later use in our applicative framework the fluctuations of $\hat{\lambda}_{\mathcal K(i)+\ell}$ when $\omega_i$ does {\it not} satisfy the separation condition, and when $X$ is a standard Gaussian matrix. For this, we have the following theorem \cite{JOH00,FEL10,BAI05}.
\begin{theorem}
	\label{th:TW}
	Let $X$ be standard Gaussian, then if $0<\omega_i<\sqrt{c}$,
			\begin{align*}
				N^{\frac23}\frac{\hat{\lambda}_{\mathcal K(i)+\ell}-(1+\sqrt{c})^2}{(1+\sqrt{c})^{\frac43}\sqrt{c}}\Rightarrow T_2
			\end{align*}
			and, if $-\sqrt{c}<\omega_i<0$,
			\begin{align*}
				N^{\frac23}\frac{\hat{\lambda}_{\mathcal K(i)+\ell}-(1-\sqrt{c})^2}{-(1-\sqrt{c})^{\frac43}\sqrt{c}}\Rightarrow T_2
			\end{align*}
			for $\ell=1,\ldots,j_i$, as $n\to \infty$, where $T_2$ is the complex Tracy-Widom distribution function \cite{TRA96}.
\end{theorem}

The tools used to derive Theorem \ref{th:TW} are much different from those exploited here and will not be discussed. Note that in \cite{ELK07}, an extension to the case where $X$ may be correlated is provided but only considers the fluctuations of the largest eigenvalue. Similar to \cite{BIA10}, Theorem \ref{th:TW} will be used to derive tests to decide on the presence of eigenvalues outside the support of the Mar\u{c}enko-Pastur law. For failure detection purposes in sensor networks, this will be used to declare a failure prior to diagnose the fault. Then, to diagnose a failure, second order statistics of both eigenvalue and eigenspace projections when the separation property arises are needed. This is the aim of the remainder of the section. 

We now turn to the second order analysis of the eigenspectrum of $\Sigma\Sigma^\ast$ when $\omega_i$ satisfies the separation condition, and when $X$ is only assumed to satisfy {\bf A1}-{\bf A3}. We first need the following additional assumption: 
\begin{description}
\item[{\bf A4}] 
For all $z \in \CC \setminus [a,b]$, 
\begin{equation*}
\sqrt{N}\left(\alpha(z) - m(z) \right) \probto 0 
\end{equation*}
as $n\to\infty$.
\end{description} 
In practice, this assumption means that the fluctuations of the spectral measure of $XX^\ast$ are negligible with respect to those of the $\hat\lambda_k$ and $b_1^* \widehat\Pi_i b_2$, which will be shown to be of order $\sqrt{N}$. This assumption is satisfied by most of the random matrix models of practical importance in our context, provided $\sqrt{N}(N/n - c) \to 0$. 
The classical illustrating example in this regard concerns the standard Gaussian case. Denote by $m_n(z)$ the Stieltjes transform \eqref{eq:mp} where $c$ is replaced with $N/n$, and let $\pi_n$ the Mar\u{c}enko-Pastur law associated with $m_n(z)$. For any $z \in \CC \setminus [a,b]$, the function $f(x) = (x-z)^{-1}$ is analytic outside the support of $\pi_n$ for $n$ sufficiently large. As a consequence, Theorem 1.1 of \cite{BAI04} shows that $\sqrt{N} ( \alpha(z) - m_n(z) ) \probto 0$. Assuming in addition that $\sqrt{N}(N/n - c) \to 0$, it is not difficult to show from \eqref{eq:mp} that $\sqrt{N} ( m_n(z) - m(z) ) \to 0$.\footnote{It is useful to note that when the convergence rate of $N/n$ is slower than $1/\sqrt{N}$, our analysis remains true if we replace $m(z)$ with a finite horizon deterministic equivalent \cite{HAC07,HAC08}. For simplicity, we have chosen not to enter into these details here.}

For practical purposes, we shall also assume: 
\begin{description}
\item[{\bf A5}] 
Each $\omega_i$, $1\leq i\leq t$, satisfies the separation condition.
\end{description} 

The main result of this section is the following theorem.
\begin{theorem}
\label{th:2ord} 
Assume {\bf A1}-{\bf A5}. For $i=1,\ldots,t$, let 
\begin{equation*}
V_{i,n} = \sqrt{N} U_i^* \left( \widehat\Pi_i - \zeta_i I_N \right) U_i 
\end{equation*}
and 
\begin{equation*}
L_{i,n} = \sqrt{N} \begin{bmatrix} 
\hat\lambda_{{\mathcal K}(i)+1} - \rho_i \\
\vdots \\
\hat\lambda_{{\mathcal K}(i)+j_i} - \rho_i 
\end{bmatrix} . 
\end{equation*}
For $\rho \in \RR \setminus [a,b]$, let  
\[
D(\rho) = 
\begin{bmatrix} 
\frac{h(\rho)(1+h(\rho)) h''(\rho)}
{h'(\rho)^3}  & 
- \frac{h(\rho)(1+h(\rho))}{h'(\rho)^2}  \\
- \frac{\rho}{h'(\rho)} & 0 
\end{bmatrix}  
\]
and 
\[
R(\rho) = \begin{bmatrix}
m'(\rho) - m(\rho)^2 & m''(\rho)/2 - m(\rho) m'(\rho) \\
m''(\rho)/2 - m(\rho) m'(\rho) & 
m^{(3)}(\rho)/6 - m'(\rho)^2  
\end{bmatrix}  
\]
where $m^{(3)}$ is the third derivative of $m$. Consider the matrices 
\[
\begin{bmatrix} G_{i} \\ K_{i} \end{bmatrix} = 
\left( \left(D(\rho_i) R(\rho_i) D(\rho_i)^*\right)^{\frac12} 
\otimes I_{j_i} \right) 
\begin{bmatrix} M_{1,i} \\ M_{2,i} \end{bmatrix} 
\]
where $M_{1,1}, M_{2,1}, \ldots, M_{1,t}, M_{2,t}$ are independent GUE matrices such that $M_{1,i}$ and $M_{2,i}$ are both $j_i \times j_i$ matrices.\footnote{We remind that matrices from the Gaussian unitary ensemble, or GUE matrices, are random Hermitian matrices with independent standard real Gaussian diagonal entries and independent standard complex Gaussian upper-diagonal entries \cite{MUI82}.} Let $L_i$ be the $\RR^{j_i}$-valued vector of eigenvalues of $K_i$ arranged in decreasing order. Then\footnote{It is clear that if {\bf A5} is not fulfilled, Theorem \ref{th:2ord} generalizes by letting the index $i$ in \eqref{eq:2ord} span only the subset $\{1,\ldots,p,q,\ldots,t\}$.}
\begin{equation}
\label{eq:2ord} 
\left( \left( V_{i,n}, L_{i,n} \right) \right)_{i=1}^t
\Rightarrow
\left( \left( G_{i} , L_{i} \right) \right)_{i=1}^t. 
\end{equation}
\end{theorem}
\begin{figure}
  \centering
  \begin{tikzpicture}[font=\footnotesize,scale=1]
    \renewcommand{\axisdefaulttryminticks}{4} 
    \tikzstyle{every axis y label}+=[yshift=-10pt] 
    \tikzstyle{every axis x label}+=[yshift=5pt]
    \pgfplotsset{every axis legend/.append style={cells={anchor=west},fill=white, at={(0.98,0.98)}, anchor=north east, font=\scriptsize }}
    \begin{axis}[
      xmin=-1.5,
      ymin=0,
      xmax=1.5,
      ymax=1.9,
      bar width=2pt,
      grid=major,
      ymajorgrids=false,
      scaled ticks=true,
      xlabel={Centered-scaled projection $|\hat{u}_1^\ast u_1|^2$},
      ylabel={Density}
      ]
      \addplot+[ybar,mark=none,color=black,fill=blue!40!white] coordinates{
(-1.500,0.000)(-1.425,0.000)(-1.350,0.000)(-1.275,0.000)(-1.200,0.001)(-1.125,0.001)(-1.050,0.007)(-0.975,0.016)(-0.900,0.019)(-0.825,0.045)(-0.750,0.065)(-0.675,0.131)(-0.600,0.197)(-0.525,0.323)(-0.450,0.433)(-0.375,0.625)(-0.300,0.775)(-0.225,1.061)(-0.150,1.220)(-0.075,1.385)(0.000,1.444)(0.075,1.337)(0.150,1.277)(0.225,0.951)(0.300,0.779)(0.375,0.543)(0.450,0.328)(0.525,0.193)(0.600,0.108)(0.675,0.044)(0.750,0.016)(0.825,0.008)(0.900,0.000)(0.975,0.000)(1.050,0.000)(1.125,0.000)(1.200,0.000)(1.275,0.000)(1.350,0.000)(1.425,0.000)(1.500,0.000)
      };
      \addplot[smooth,red,line width=0.5pt] plot coordinates{
(-1.500,0.000)(-1.425,0.000)(-1.350,0.000)(-1.275,0.000)(-1.200,0.000)(-1.125,0.000)(-1.050,0.001)(-0.975,0.003)(-0.900,0.008)(-0.825,0.019)(-0.750,0.040)(-0.675,0.079)(-0.600,0.145)(-0.525,0.247)(-0.450,0.393)(-0.375,0.582)(-0.300,0.803)(-0.225,1.031)(-0.150,1.232)(-0.075,1.372)(0.000,1.422)(0.075,1.372)(0.150,1.232)(0.225,1.031)(0.300,0.803)(0.375,0.582)(0.450,0.393)(0.525,0.247)(0.600,0.145)(0.675,0.079)(0.750,0.040)(0.825,0.019)(0.900,0.008)(0.975,0.003)(1.050,0.001)(1.125,0.000)(1.200,0.000)(1.275,0.000)(1.350,0.000)(1.425,0.000)(1.500,0.000)
};
\legend{{Histogram of $\sqrt{N}(|\hat{u}_1^\ast u_1|^2-\zeta_1)$},{Gaussian $\mathcal N(0,C(\rho_1)_{1,1})$}}
    \end{axis}
  \end{tikzpicture}
  \begin{tikzpicture}[font=\footnotesize,scale=1]
    \renewcommand{\axisdefaulttryminticks}{4} 
    \tikzstyle{every axis y label}+=[yshift=-10pt] 
    \tikzstyle{every axis x label}+=[yshift=5pt]
    \pgfplotsset{every axis legend/.append style={cells={anchor=west},fill=white, at={(0.98,0.98)}, anchor=north east, font=\scriptsize }}
    \begin{axis}[
      xmin=-5,
      ymin=0,
      xmax=5,
      ymax=0.5,
      bar width=2pt,
      grid=major,
      ymajorgrids=false,
      scaled ticks=true,
      xlabel={Centered-scaled projection $|\hat{u}_1^\ast u_1|^2$},
      ylabel={Density}
      ]
      \addplot+[ybar,mark=none,color=black,fill=blue!40!white] coordinates{
(-5.000,0.001)(-4.750,0.005)(-4.500,0.004)(-4.250,0.005)(-4.000,0.005)(-3.750,0.007)(-3.500,0.012)(-3.250,0.015)(-3.000,0.026)(-2.750,0.033)(-2.500,0.040)(-2.250,0.052)(-2.000,0.072)(-1.750,0.103)(-1.500,0.124)(-1.250,0.150)(-1.000,0.204)(-0.750,0.243)(-0.500,0.314)(-0.250,0.358)(0.000,0.387)(0.250,0.387)(0.500,0.375)(0.750,0.333)(1.000,0.252)(1.250,0.214)(1.500,0.125)(1.750,0.080)(2.000,0.042)(2.250,0.020)(2.500,0.005)(2.750,0.001)(3.000,0.000)(3.250,0.000)(3.500,0.000)(3.750,0.000)(4.000,0.000)(4.250,0.000)(4.500,0.000)(4.750,0.000)(5.000,0.000)
      };
      \addplot[smooth,red,line width=0.5pt] plot coordinates{
(-5.000,0.000)(-4.750,0.000)(-4.500,0.000)(-4.250,0.000)(-4.000,0.001)(-3.750,0.001)(-3.500,0.003)(-3.250,0.005)(-3.000,0.009)(-2.750,0.017)(-2.500,0.029)(-2.250,0.046)(-2.000,0.071)(-1.750,0.104)(-1.500,0.145)(-1.250,0.191)(-1.000,0.239)(-0.750,0.286)(-0.500,0.324)(-0.250,0.350)(0.000,0.358)(0.250,0.350)(0.500,0.324)(0.750,0.286)(1.000,0.239)(1.250,0.191)(1.500,0.145)(1.750,0.104)(2.000,0.071)(2.250,0.046)(2.500,0.029)(2.750,0.017)(3.000,0.009)(3.250,0.005)(3.500,0.003)(3.750,0.001)(4.000,0.001)(4.250,0.000)(4.500,0.000)(4.750,0.000)(5.000,0.000)
};
\legend{{Histogram of $\sqrt{N}(|\hat{u}_1^\ast u_1|^2-\zeta_1)$},{Gaussian $\mathcal N(0,C(\rho_1)_{1,1})$}}
    \end{axis}
  \end{tikzpicture}
  \caption{Empirical and theoretical distribution of the fluctuations of $\hat{u}_1$ with $r=1$, $X$ has i.i.d. zero mean variance $1/n$ entries, $N/n=1/8$, $N=256$ and $\omega_1=1$ (top) or $\omega_1=0.5$ (bottom).}
  \label{fig:fluct_eigenvector}
\end{figure}
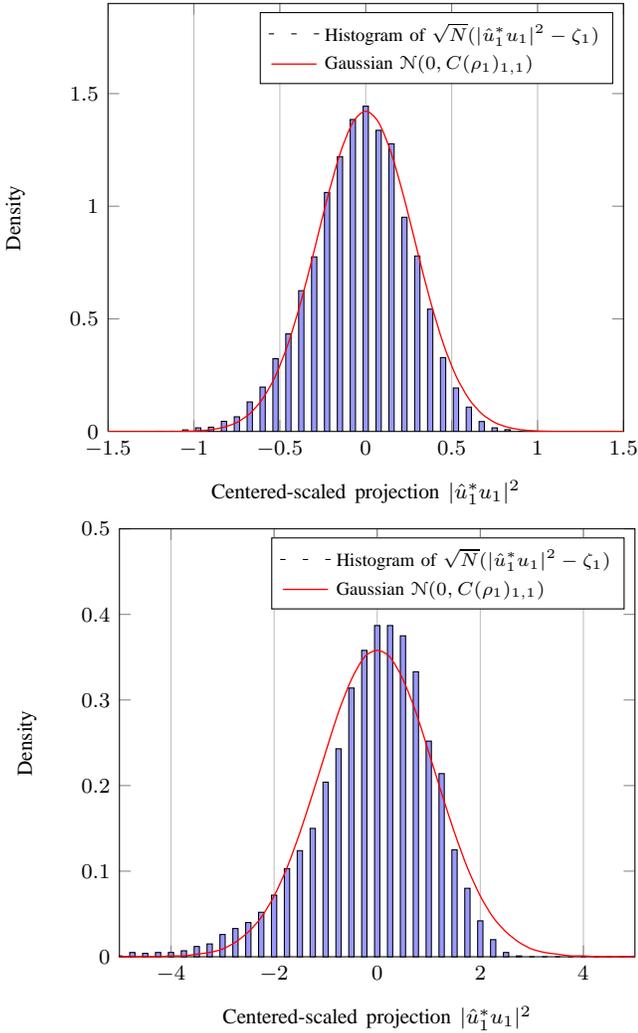

\begin{remark}
To be more precise, $\left( V_{i,n}, L_{i,n} \right)$ can be cast into an $\RR^{j_i^2+j_i}$-valued random vector after rearranging the real and imaginary parts of the elements of $V_{i,n}$ and taking into account the Hermitian symmetry constraint. Hence, the convergence in law specified by \eqref{eq:2ord} acts on the space of probability measures on $\RR^{r+j_1^2+\cdots+j_t^2}$. 
\end{remark} 

Theorem \ref{th:2ord} provides a very general expression of the joint limiting fluctuations of both eigenvalues and eigenspace projections. It is particularly interesting to note that the fluctuations of $(V_{i,n},L_{i,n})$ are asymptotically independent across $i$. 

This theorem is an interesting new result to the field of large dimensional random matrix theory, although it might be difficult to use in practice since one needs to derive explicitly the joint density of $(G_i,L_i)_{i=1}^t$. Nonetheless, Theorem \ref{th:2ord} can be immediately put to practice in two scenarios. The first scenario corresponds to the case where only the fluctuations of the eigenspace projection vector $(V_{i,n})_{i=1}^t$ is of interest. In this case, $(G_i)_{i=1}^t$ is a correlated Gaussian random variable. Reminding that $M_{1,i}$ can be seen as an $\RR^{j_i^2}$-valued Gaussian vector with entries of zero mean and unit variance, $(G_i)_{i=1}^t$ can then be seen as a random Gaussian vector in $\RR^{j_1^2+\ldots+j_t^2}$ with entries $(j_{i-1}^2-1)$ to $j_i^2$ of zero mean and variance $\left[D(\rho_i)R(\rho_i)D(\rho_i)^\ast\right]_{11}$, for each $i\in\{1,\ldots,t\}$ with $j_0=0$. The second scenario of interest, which is discussed at length in the following, corresponds to the case where the multiplicities of the eigenvalues of $P$ are all equal to one. In this case, $M_{1,i}$ and $M_{2,i}$ are independent Gaussian variables and we immediately have the following corollary: 
\begin{corollary}
	\label{co:2ord}
Consider the setting of Theorem \ref{th:2ord}. Assume in addition that $j_i = 1$ for all $i$. Then
\begin{equation*} 
\left( \left( V_{i,n}, L_{i,n} \right) \right)_{i=1,\ldots,r} \Rightarrow {\mathcal N}(0, R)
\end{equation*}
with 
\begin{equation*}
R = \begin{bmatrix} 
D(\rho_1) R(\rho_1) D(\rho_1)^* \\ 
& \ddots \\ & & D(\rho_r) R(\rho_r) D(\rho_r)^* 
\end{bmatrix} . 
\end{equation*}
\end{corollary}

After some calculus, in the standard Gaussian case, we further have:
\begin{corollary}
	\label{co-bf2}
	Under the assumptions of Corollary \ref{co:2ord}, if $X$ is a standard Gaussian matrix, then $D(\rho_i)R(\rho_i)D(\rho_i)^\ast=C(\rho_i)$, where
\begin{equation*}
	C(\rho_{i}) \triangleq
\begin{bmatrix}
	\frac{c^2(1+\omega_i)^2}{(c+\omega_i)^2(\omega_i^2-c)}\left(c\frac{(1+\omega_i)^2}{(c+\omega_i)^2} + 1 \right) & \frac{(1+\omega_i)^3c^2}{(\omega_i+c)^2\omega_i} \\
	\frac{(1+\omega_i)^3c^2}{(\omega_i+c)^2\omega_i} & \frac{c(1+\omega_i)^2(\omega_i^2-c)}{\omega_i^2}	\end{bmatrix}.
\end{equation*} 
\end{corollary}

Due to its simple expression, Corollary \ref{co-bf2} is particularly handy to use in the context of failure diagnosis when hypothetical failures are characterized by distinct values of $\omega_i$, as will be shown in Section \ref{sec:application}.

In Figure \ref{fig:fluct_eigenvector}, the histogram of a simulation of $10\,000$ realizations of the projection $V_{1,n}=\sqrt{N}(|\hat{u}_1^\ast u_1|^2-\zeta_1)$, with $u_1=U_1\in\CC^{N}$, $\hat{u}_1\hat{u}_1^\ast=\widehat{\Pi}_1$ of unit rank, and $X$ standard Gaussian, is depicted against the asymptotic Gaussian law derived in Corollary \ref{co-bf2}, for $r=1$, $N/n=1/8$, $N=256$ and $\omega_1$ successively equal to $1$ and $0.5$. In this scenario, $\sqrt{N/n}\simeq 0.35$. For $\omega_1=1$, the simulation shows a rather accurate fit between asymptotic theory and simulation. For $\omega_1=0.5$, the Gaussian approximation is much less accurate. This is due to the value $\omega_1=0.5$, which is rather close to $\sqrt{N/n}\simeq 0.35$. The value $N=256$ is here insufficient for the large dimensional behaviour of the fluctuations of $|\hat{u}_1^\ast u_1|^2$ to arise. This behaviour will have important consequences for the question of diagnosing failures which are difficult to observe.

The remainder of this section is devoted to the proof of Theorem \ref{th:2ord}. We start with the following lemma, which deals with the asymptotic behavior of the $V_{i,n}$. This lemma will be proved in Appendix \ref{anx:cltvep}.
\begin{lemma}
\label{lm:cltvep} 
Assume {\bf A1}-{\bf A5}. Let 
\begin{align*}
\bar V_{i,n} &= 
\sqrt{N} \left( 
\frac{h(\rho_i)(1+h(\rho_i)) h''(\rho_i)}
{h'(\rho_i)^3} 
U_i^* (Q(\rho_i) - m(\rho_i)I) U_i \right. \\ 
&\left.  - 
\frac{h(\rho_i)(1+h(\rho_i))}{h'(\rho_i)^2} 
U_i^* (Q'(\rho_i) - m'(\rho_i)I) U_i \right) . 
\end{align*} 
Then for any $i\in \{1, \ldots, t \}$, 
\begin{equation*}
V_{i,n} - \bar V_{i,n} \probto 0 . 
\end{equation*}
\end{lemma} 

We now consider the isolated extreme eigenvalues. In order to study the asymptotic behavior of these eigenvalues, we shall adapt to our situation the approach of \cite{ben-gui-mai-11}. For $i=1,\ldots, t$, consider real numbers $x_1(i) > y_1(i) > x_2(i) > y_2(i) > \cdots > y_{j_i}(i)$. Since the separation condition is satisfied by assumption for each $i = 1,\ldots,t$, the equation $\det \widehat H(x)=0$ has $r$ roots outside $[a,b]$ with probability one for all $n$ large. Therefore, we have the equivalence relation
\begin{gather} 
\left( 
 x_\ell(i)  > \sqrt{N} ( \hat\lambda_{{\mathcal K}(i) + \ell} 
- \rho_i ) > y_\ell(i), \right. 
\ \ \ \ \ \ \ \ \ \ \ \ \ 
\nonumber \\
\ \ \ \ \ \ \ \ \ \ \ \ \ 
\left. \phantom{\hat\lambda_{{\mathcal K}(i) + \ell}} 
i = 1,\ldots, t, \quad \ell=1, \ldots, j_i 
\right) \nonumber \\ 
\Leftrightarrow \nonumber \\ 
\left( N^{j_i}  
\det\widehat H\Bigl(\rho_{i} + \frac{x_\ell(i)}{\sqrt{N}} \Bigr) \, 
\det\widehat H\Bigl(\rho_{i} + \frac{y_\ell(i)}{\sqrt{N}} \Bigr) < 0, 
\right. \nonumber \\
\ \ \ \ \ \ \ \ \ \ \ \ \ 
\left. \phantom{\frac 1N} 
\quad i = 1,\ldots, t, \quad \ell=1, \ldots, j_i 
\right) . 
\label{equiv-vap}
\end{gather} 
This equivalence leads us to study the limits of the finite dimensional distributions of the $\RR^t$-valued random process $ \left[ N^{j_i/2} \det \widehat H(\rho_i +  x/\sqrt{N}) \right]_{i=1}^t$ in the parameter $x$. This is given by the following lemma, proved in Appendix \ref{anx:vap}. 
\begin{lemma}
\label{lm:vap} 
Assume {\bf A1}-{\bf A5}. 
Define the $\RR^t$-valued random process 
\begin{align*} 
&\chi_n(x) = \\
&\Bigl[ N^{j_i/2} \det \widehat H(\rho_i +  x/\sqrt{N}) - \Bigl( \prod_{\ell\neq i} (1 - \beta_\ell / \beta_i)^{j_\ell} \Bigr) \\ 
&\times \det\left( \sqrt{N} \beta_i \rho_i U_i^* ( Q(\rho_i)  - m(\rho_i) I) U_i + \beta_i h'(\rho_i)  x I \right)\Bigr]_{i=1}^t
\end{align*}
where $\beta_i = \omega_i / (1+\omega_i)$. Then 
\begin{equation*}
	(\chi_n(x_1), \ldots, \chi_n(x_p)) \probto 0
\end{equation*}
for every finite sequence $(x_1, \ldots, x_p)$. 
\end{lemma}

Let $B$ be a rectangle of $\RR^{j_1^2+\cdots+j_t^2}$. The real and imaginary parts of the elements of the $V_{i,n}$ defined in Theorem \ref{th:2ord} can be stacked into a $\RR^{j_1^2+\cdots+j_t^2}$-valued vector $V_n$ when taking into account the Hermitian symmetry constraint. A vector $\bar V_n$ with the same size can be constructed similarly from $\bar V_{i,n}$ defined in Lemma \ref{lm:cltvep}. Let $L_n$ be the $\RR^r$-valued vector $L_n = [ L_{1,n}^\trans,\ldots, L_{t,n}^\trans ]^\trans$ (see Theorem \ref{th:2ord}) and let $C$ be the rectangle of $\RR^r$ determined by the left hand side of \eqref{equiv-vap}, so that this event is written $[ L_n \in C ]$.
Let $\bar L_n$ be the vector obtained by arranging the eigenvalues of 
the matrices 
\begin{equation*}
\bar T_{i,n} = 
- \frac{\rho_i}{h'(\rho_i)} 
\sqrt{N} U_i^* ( Q(\rho_i)  - m(\rho_i) I) U_i
\end{equation*}
for $i=1,\ldots,t$, similarly to the elements of $L_n$. From Lemma \ref{lm:cltvep}, Lemma \ref{lm:vap}, and the discussion preceding Lemma \ref{lm:vap}, we have 
\begin{equation*}
\PP\left[ V_n \in B, L_n \in C \right] - \PP\left[ \bar V_n \in B, \bar L_n \in C \right] \to 0 
\end{equation*}
as $n\to\infty$, for arbitrary rectangles $B$ and arbitrary rectangles $C$ specified at the left hand side of \eqref{equiv-vap}. Observe that 
\begin{equation*}
\begin{bmatrix} \bar V_{i,n} \\ \bar T_{i,n} \end{bmatrix} = 
\sqrt{N} \left( D(\rho_i) \otimes I_{j_i} \right) 
\begin{bmatrix} 
U_i^* ( Q(\rho_i)  - m(\rho_i) I) U_i \\
U_i^* ( Q'(\rho_i)  - m'(\rho_i) I) U_i
\end{bmatrix} .
\end{equation*}

In order to terminate the proof of Theorem \ref{th:2ord}, we shall make use of the following lemma, that we state in a slightly more general form than needed here. 

\begin{lemma}
\label{lm:tlc} 
Assume {\bf A1}-{\bf A4}. 
Let $f_1, \ldots, f_t$ and $g_1, \ldots, g_t$ be real functions analytical on a neighborhood of $[a,b]$. Let $S_n$ be the $t$-uple of random matrices 
\begin{equation*}
S_n = \begin{pmatrix} \sqrt{N} \begin{bmatrix} 
U_i^* f_i(XX^*) U_i - \left( \int f_i(\lambda) d\pi(\lambda) \right) I_{j_i} \\
U_i^* g_i(XX^*) U_i - \left( \int g_i(\lambda) d\pi(\lambda) \right) I_{j_i} 
\end{bmatrix}  
\end{pmatrix}_{i=1}^t.  
\end{equation*}
For $i=1,\ldots, t$, define the covariance matrices 
\begin{equation*}
R_i = \! \int \!\! 
\left( 
\begin{bmatrix} 
f_i - \int f_i d\pi \\
g_i - \int g_i d\pi 
\end{bmatrix}
\begin{bmatrix} 
(f_i - \int f_i d\pi) & (g_i - \int g_i d\pi) 
\end{bmatrix}
\right) 
d\pi . 
\end{equation*}
Then $S_n$ converges in distribution towards 
\begin{equation}
\label{clt-gue} 
\begin{pmatrix} 
\left( R_i^{\frac12} \otimes I_{j_i} \right) 
\begin{bmatrix} M_{1,i} \\ M_{2,i} \end{bmatrix} 
\end{pmatrix}_{i=1}^t  
\end{equation} 
where the matrices $(M_{1,1}, M_{2,1}, \ldots, M_{1,t}, M_{2,t})$ are independent GUE matrices such that $M_{1,i}$ and $M_{2,i}$ have dimensions $j_i \times j_i$. 
\end{lemma} 

The proof is provided in Appendix \ref{anx:tlc}.

Applying this lemma with $f_i(\lambda) = 1/(\lambda-\rho_i)$ and $g_i(\lambda) = 1/(\lambda-\rho_i)^2$, $R_i$ takes the value $R(\rho_i)$ provided in the statement of Theorem \ref{th:2ord}. It results that 
\begin{equation*}
\left( 
\begin{bmatrix} \bar V_{i,n} \\ \bar T_{i,n} \end{bmatrix}
\right)_{i=1}^t
\Rightarrow
\left( 
\begin{bmatrix} G_{i} \\ K_{i} \end{bmatrix}
\right)_{i=1}^t 
\end{equation*}
where the convergence takes place on the space of probability measures on $\RR^{2(j_1^2+\cdots+j_t^2)}$. This completes the proof of Theorem \ref{th:2ord}.

\section{Application}
 \label{sec:application}

 In this section, we provide a general framework for local failure detection and diagnosis in large sensor networks, such as the examples proposed in Section \ref{sec:failure}, based on the results of Section \ref{sec:main_res}. This framework is a two-step approach for successively (i) detecting failures within a given maximally acceptable false alarm rate and (ii) upon positive detection, diagnosing the failures with high probability. Simulations are then run to validate the proposed algorithms.

 We recall that we assume a number $K$ of failure scenarios indexed by $1\leq k\leq K$. Scenario $k$ is characterized by the population covariance matrix $I_N+P_k$. We additionally denote $P_0=0$ for convenience. We consider precisely the detection and localization to be made for failure models of the type $\Sigma=(I_N+P_k)^\oh X$, $1\leq k\leq K$, with $P_k=\sum_{i=1}^{t_k} \omega_{k,i}U_{k,i}U_{k,i}^\ast$ of rank $r_k=\sum_{i=1}^{t_k}j_{k,i}$, where $U_{k,i}\in\CC^{N\times j_{k,i}}$ and $\omega_{k,1}>\ldots>\omega_{k,s_k}>0>\omega_{k,s_k+1}>\ldots>\omega_{k,t_k}$, and $X$ standard Gaussian. The eigenvalues of $\Sigma\Sigma^\ast$ are denoted and ordered as before by $\hat{\lambda}_1\geq\ldots\geq\hat{\lambda}_N$. We denote $\mathcal H_0$ the null hypothesis for the model $\Sigma=X$ and $\mathcal H_k$ the hypothesis for a failure of type $k$. For the failure scenario $k$, we also take $p_k$ to be zero or the smallest index $i$ such that $\omega_{k,i}>\sqrt{c}$, and $q_k$ to be $t_k+1$ or the largest index $i$ such that $\omega_{k,i}<-\sqrt{c}$.

\begin{remark}
	This model (and in particular the models of Section \ref{sec:ex1} and Section \ref{sec:ex2}) may be extended by introducing deterministic linear time correlation between the successive observations $s_1,\ldots,s_n$. That is, we can write $X=\tilde{X}T^\oh$ for some time-correlation matrix $T$ and a standard Gaussian matrix $\tilde{X}$. In this scenario, $X$ being left-unitarily invariant, the localization scheme proposed extends naturally to this scenario, as long as the eigenvalue distribution of $T$ converges weakly to a compactly supported distribution as $N$ grows large and that $T$ has asymptotically no eigenvalue outside this support. Alternatively, a whitening procedure may be used prior to failure detection using $\Sigma'=\Sigma T^{-\oh}$ as the random matrix under study, provided that $T$ is invertible.
\end{remark}

 \subsection{Detection algorithm}
 \label{sec:detect_algo}
 As stated in the introduction, the detection phase relies on existing results, and more specifically on the fluctuations of the largest eigenvalues given by Theorem \ref{th:TW}.
 The detection algorithms proposed here parallel that introduced in \cite{BIA10} in the context of collaborative signal sensing. The objective is to decide between hypothesis $\mathcal H_0$ and its complementary $\bar{\mathcal H}_0$. 
 
 First assume that all $P_k$ only have non-negative eigenvalues. From Theorem \ref{th-spk}, the largest eigenvalue $\hat{\lambda}_1$ of $\Sigma\Sigma^\ast$ tends to the right edge of the support $S_\pi$ of the Mar\u{c}enko-Pastur law for all large $n$ under $\mathcal H_0$, while $\hat{\lambda}_1$ is found away from this edge under $\bar{\mathcal H}_0$ {\it if} the largest eigenvalue $\omega_{k,1}$ of $P_k$ exceeds $\sqrt{c}$. We will therefore assume in the following that $\omega_{k,1}>\sqrt{c}$ is verified for all $k$. That is, we assume that $c_N\triangleq N/n<c_+$ where $c_+$ is defined as
 \begin{equation*}
	 c_+ \triangleq \inf \left\{ \omega_{k,1}^2,~1\leq k\leq K \right\}.
 \end{equation*} 
 
 This condition allows for a theoretically almost sure error detection, as $N,n\to\infty$. We then rely on Theorem \ref{th:TW} to design an appropriate hypothesis test. Our test consists in rejecting hypothesis $\mathcal H_0$ if the probability in favor of $\mathcal H_0$ is sufficiently low. That is, for a given acceptable {\it false alarm rate} $\eta$,\footnote{We recall that the false alarm rate is the probability of declaring $\bar{\mathcal H}_0$ under true hypothesis $\mathcal H_0$.} the statistical test is defined as
 \begin{equation}
	 \label{eq:test+}
	 \hat{\lambda}'_1 \overset{{\mathcal H}_0}{\underset{\bar{\mathcal H}_0}{\lessgtr}} (T_2)^{-1}(1-\eta)
 \end{equation}
 where $\hat{\lambda}'_1$ is given by
 \begin{equation*}
	 \hat{\lambda}'_1 \triangleq N^{\frac23} \frac{\hat{\lambda}_1-(1+\sqrt{c_N})^2}{(1+\sqrt{c_N})^{\frac43}c_N^\oh}.
 \end{equation*}
 That is, the test verifies whether $\hat{\lambda}_1$ exceeds some threshold above which the probability for $\mathcal H_0$ is less than $\eta$.

 If the matrices $P_k$ are now all non-positive definite, then, symmetrically, we need to set $c_N$ such that the smallest eigenvalue $\hat{\lambda}_N$ of $\Sigma\Sigma^\ast$ is visible on the left-hand side of $S_\pi$. That is, we take $N,n$ to be such that $c_N<c_-$, with $c_-$ defined as
 \begin{equation*} 
	 c_- \triangleq \inf \left\{ \omega_{k,t_k}^2,~1\leq k\leq K\right\}.
 \end{equation*}
The decision test is in that case given by
 \begin{equation}
	 \label{eq:test-}
	 \hat{\lambda}'_N \overset{{\mathcal H}_0}{\underset{\bar{\mathcal H}_0}{\lessgtr}} (T_2)^{-1}(1-\eta)
 \end{equation}
 where $\hat{\lambda}'_N$ is defined as
 \begin{equation*}
	 \hat{\lambda}'_N \triangleq N^{\frac23} \frac{\hat{\lambda}_N-(1-\sqrt{c_N})^2}{-(1-\sqrt{c_N})^{\frac43}c_N^\oh}.
 \end{equation*}

 The above test is particularly suited to the model of Section \ref{sec:ex2} in which the matrices $P_k$ are non-positive definite when $\mu_k=0$ and $\alpha_k=-1$ for all $k$, corresponding to a sudden drop of a zero mean random parameter $\theta(k)$ to zero.

 When the matrices $P_k$ have both positive and negative eigenvalues, a deterministic choice has to be made by the experimenter. In the most general setting, to ensure a false alarm rate lower than $\eta$, one has to choose two scalars $a(\eta),b(\eta)\in \RR\cup \{-\infty,\infty\}$ such that
 \begin{equation}
	 \label{eq:Pab}
	 P\left(\left\{\hat{\lambda}'_1>a(\eta)\right\} \cup \left\{\hat{\lambda}'_N>b(\eta)\right\}\right) \leq \eta.
 \end{equation}
 The choice of $a(\eta),b(\eta)$ depends primarily on the structure of $P_k$ and will impact the correct detection rate for fixed false alarm rates.

In \cite{BIA08}, the asymptotic independence of the fluctuations of the largest and the smallest eigenvalues of GUE matrices is proved, while the same result for the eigenvalues $\hat{\lambda}_1$ and $\hat{\lambda}_N$ of $\Sigma\Sigma^\ast$ under $\mathcal H_0$ is conjectured. Following this conjecture, \eqref{eq:Pab} would become asymptotically
 \begin{equation*}
T_2(a(\eta))T_2(b(\eta))>1-\eta.
 \end{equation*}
For any fixed $b$, taking $b(\eta)=b$, the hypothesis test now becomes 
\begin{align*}
	\min\left\{ b-\hat{\lambda}'_N,~(T_2)^{-1}\left( \frac{1-\eta}{T_2(b)}\right) - \hat{\lambda}'_1 \right\} \overset{\bar{\mathcal H}_0}{\underset{\mathcal H_0}{\lessgtr}} 0.
\end{align*}
In particular, for $b=\infty$, $T_2(b)=1$ and then the test reduces to
\begin{align*}
	 \hat{\lambda}'_1 \overset{{\mathcal H}_0}{\underset{\bar{\mathcal H}_0}{\lessgtr}} (T_2)^{-1}\left(1-\eta\right)
\end{align*}
which is the same test as proposed in \eqref{eq:test+}. Taking instead $a(\eta)=-\infty$, we obtain the test \eqref{eq:test-}.

For rather symmetrical distributions of the eigenvalues $\omega_{k,i}$ of $P_k$ around zero, it may be interesting to set $b(\eta)=a(\eta)$, in which case
\begin{equation*}
	b(\eta)=(T_2)^{-1}\left(\sqrt{1-\eta}\right).
\end{equation*}
In this setting, the decision test is now
\begin{align*}
	\max\left\{ \hat{\lambda}'_N,\hat{\lambda}'_1 \right\} \overset{{\mathcal H}_0}{\underset{\bar{\mathcal H}_0}{\lessgtr}} (T_2)^{-1}\left(\sqrt{1-\eta}\right).
\end{align*}

In the following section, we assume that the procedure of failure detection was achieved successfully and that we are now interested in localizing the failures.

 \subsection{Localization algorithm}
 \label{sec:localize_algo}
 We now wish to detect all possible failure events from a set of failures indexed by $k\in\{1,\ldots,K\}$. The index set $\{1,\ldots,K\}$ may gather all events accounting for a single, as well as multiple, local failures. 
 Similar to the previous sections, we denote $\rho_{k,i}=1+\omega_{k,i}+c\frac{1+\omega_{k,i}}{\omega_{k,i}}$ and $\zeta_{k,i}=\frac{1-c\omega_{k,i}^{-2}}{1+c\omega_{k,i}^{-1}}$, we define the mapping $\mathcal K_k$ to be such that $\mathcal K_k(i)=j_{k,1}+\ldots+j_{k,i-1}$ if $1\leq i\leq s_k$ and $\mathcal K_k(i)=N-(j_{k,i}+\ldots+j_{k,t_k})$ if $s_k+1\leq i\leq t_k$. Finally, we denote $\widehat{\Pi}_{k,i}$ any projector on the subspace generated by the eigenvalues $\hat\lambda_{\mathcal K_k(i)+1},\ldots,\hat\lambda_{\mathcal K_k(i)+j_{k,i}}$.

 An initial hypothesis rejection may be performed at this stage to select only those hypotheses $\mathcal H_k$ such that the $\rho_{k,i}$ are consistent with the observations $\hat{\lambda}_1,\ldots,\hat{\lambda}_N$. For instance, if the largest eigenvalue $\hat{\lambda}_1$ is significant in the system model, one may preselect from the set $\{1,\ldots,K\}$ the $L$ hypothesis indexes defined by
 \begin{equation*}
	 \arg\min_{(k_1,\ldots,k_L)\subset \{1,\ldots,K\}} \max_{k_1,\ldots,k_L} |\hat{\lambda}_1-\rho_{k_j,1}|.
 \end{equation*}

 However, if $K$ is large, many hypotheses may have very close parameters $\rho_{k,i}$, so that it is hazardous to conclude on the most likely hypothesis based only on the eigenvalues of $\Sigma\Sigma^\ast$. However, since different $P_k$ matrices have in general very distinct eigenspaces, we propose the following subspace localization test, which decides on the hypothesis $\mathcal H_{k^\star}$ for which $k^\star$ is given by
 \begin{align}
	 \label{eq:local_test}
	 k^\star=\arg\max_{k\in S} g_k\left(\left(V^k_{i,n},L^k_{i,n}\right)_{i\in \mathcal L(p_k,q_k)} \right)
 \end{align}
with $g_k$ the actual density of the vector $\left(V^k_{i,n},L^k_{i,n}\right)_{i\in \mathcal L(p_k,q_k)}$, $\mathcal L(p_k,q_k)=\{1,\ldots,p_k,q_k,\ldots,r_k\}$, $S$ the set of (remaining) indexes $k$ such that $\mathcal L(p_k,q_k)$ is non-empty, and where
\begin{equation*}
	V^k_{i,n} \triangleq \sqrt{N} U_{k,i}^\ast\left( \widehat{\Pi}_{k,i} - \zeta_{k,i} I_{j_{k,i}} \right)U_{k,i}
\end{equation*}
and
\begin{equation*}
	L^k_{i,n} \triangleq \sqrt{N} \begin{bmatrix} \hat{\lambda}_{\mathcal K_k(i)+1} - \rho_{k,i} \\ \vdots \\ \hat{\lambda}_{\mathcal K_k(i)+j_{k,i}} - \rho_{k,i} \end{bmatrix}.
\end{equation*}
Note that we need here to specify the indexation $i\in \mathcal L(p_k,q_k)$ since we do not assume {\bf A5}.

From Theorem \ref{th:2ord}, this probability can be approximated for large $n$, which provides immediately a maximum likelihood test for the most asymptotically likely $\mathcal H_k$ hypothesis. In the particular case where the $\omega_{k,i}$ all have multiplicity one, according to Corollary \ref{co-bf2}, as $N,n$ grow large, the vectors in the test \eqref{eq:local_test} are asymptotically independent and Gaussian. We therefore substitute the test \eqref{eq:local_test} by the following test, leading to the estimator $\hat{k}$ defined as
\begin{align}
	\label{eq:kstar}
	\hat{k} = \arg\max_{k\in S} \prod_{i\in \mathcal L(p_k,q_k)} f\left( (V^k_{i,n},L^k_{i,n}) ;C(\rho_{k,i})\right)
\end{align}
where $f(x;\Omega)$, $x\in\CC^m$, $\Omega\in\CC^{m\times m}$, is the $m$-variate real normal density of zero mean and covariance $C$ at point $x$, and $C(\rho_{k,i})$ is defined similarly as in Corollary \ref{co-bf2}, with $\omega_i$ replaced by $\omega_{k,i}$. Taking a log-likelihood notation, this is explicitly
\begin{align*}
	\hat{k} = \arg\min_{k\in S} \sum_{i\in\mathcal L(p_k,q_k)}& \left[ (V^k_{i,n},L^k_{i,n})C(\rho_{k,i})^{-1}(V^k_{i,n},L^k_{i,n})^\trans \right. \\ & \left. + \log\det C(\rho_{k,i}) + 2\log(2\pi) \right].
\end{align*}
In the general case where $\omega_{k,i}$ has multiplicity $j_{k,i}$, as discussed previously, it is simpler to restrict the detection test to the eigenspace projections $(V^k_{i,n})_{i\in\mathcal L(p_k,q_k)}$ only. In this case, denoting $\bar V^k_{i,n}=(V^k_{i,n,D},\sqrt{2}V^k_{i,n,U})\in\RR^{j_{k,i}^2}$ with $V^k_{i,n,D}\in\RR^{j_{k,i}}$ the vector of the diagonal entries of $V^k_{i,n}$ and $V^k_{i,n,U}\in\RR^{(j_{k,i}^2-j_{k,i})}$ the vector of the real and imaginary parts of the upper-diagonal entries of $V^k_{i,n}$, we obtain the test
\begin{align}
	\label{eq:kstar2}
	\hat{k} = \arg\min_{k\in S} \sum_{i\in\mathcal L(p_k,q_k)}& \Big[\frac1{[C(\rho_{k,i})]_{11}} (\bar{V}^k_{i,n})^\trans{\bar{V}^k_{i,n}}\nonumber \\ &+ j_{k,i}^2 \log ([C(\rho_{k,i})]_{11}) + j_{k,i}^2\log (2\pi) \Big].
\end{align}

\begin{remark}
	We provide below some remarks and discuss the advantages of the detection tests proposed in Section \ref{sec:detect_algo} and the localization algorithms \eqref{eq:kstar}--\eqref{eq:kstar2} compared to the optimum maximum likelihood approach:
	\begin{itemize}
		\item the detection algorithms proposed in Section \ref{sec:detect_algo} are very versatile, as they adapt to multiple failure scenarios showing small rank perturbations in the population covariance, and provide a theoretical expression of the minimum ratio $c_N=N/n$ necessary for detectability;
		\item unlike the traditional maximum-likelihood approach which tests the joint distribution of $\Sigma$ for all hypotheses $1\leq k\leq K$, and therefore leads to calculus of the order $\mathcal O(N^3)$ (or $\mathcal O(N^2)$ with some simplification methods) for each $k$, the proposed localization algorithm \eqref{eq:kstar} is based on a test requiring for each $k$ (taken from a possibly reduced subset of $\{1,\ldots,K\}$) eigenvector projections of computational load of order $\mathcal O(N)$; 
		\item we may decide not to consider the joint fluctuations of all eigenvalues found outside $S_\pi$, but only some of them. This leads to an asymptotically less efficient, although much faster, algorithm, where $\mathcal L(p_k,q_k)$ in \eqref{eq:kstar} is replaced by $\mathcal L(p',q')$ for given $p'\leq p_k$, $q'\leq q_k$, for all $k$. For $N$ not too large, it is in fact preferable to consider only a few eigenvalues and eigenspace projections simultaneously, due to convergence speed limitations of the limiting normal distributions;
		\item the entries $L^k_{i,n}$ of the vector $(V^k_{i,n},L^k_{i,n})$ may also be discarded, especially in scenarios where eigenvalues of $P_k$ are very similar for each hypothesis $\mathcal H_k$. This may again increase the convergence speed of the asymptotic approximation for not-too-large $N$, while it is expected to perform worse for large $N$. 
	\end{itemize}
\end{remark}

So far, we have performed failure detection under the important assumption that the failure scenarios form a discrete set $\{1,\ldots,K\}$. This assumes in particular that the failure amplitudes are known prior to detection and localization. In the next section, we use Corollary \ref{co-bf2} to improve this approach in the particularly simple example of Section \ref{sec:ex2}, when the failure amplitude is a priori unknown.

\subsection{Extension to unknown failure amplitude}
\label{sec:unknown_failure}

In this section, we assume the scenario where the eigenvectors of the perturbation matrix $P_k$ are independent of the amplitude of the failure parameters, in the sense that a change in magnitude of the failure of type $k$ does not affect the eigenspaces of $P_k$. This is for instance the case of the single-failure scenario of Section \ref{sec:ex2}, for which we recall that $P_k$ expresses as $P_k=\beta_k R^{-\oh}He_ke_k^\ast H^\ast R^{-\oh}$ with $\beta_k$ the failure parameter. We now assume $\beta_k$ unknown, which is a more realistic assumption than assuming it perfectly known in advance. We also suppose that $X$ has i.i.d. Gaussian entries. Based on a simple extension of the algorithm presented in Section \ref{sec:localize_algo} to unknown $\omega_k$, we provide hereafter a second localization algorithm.

For notational convenience, we assume $P_k=\omega_ku_ku_k^\ast$ for each $k$ and that $\omega_k>\sqrt{c}$, unknown. We then denote $\hat{\lambda}$ the largest eigenvalue of $\Sigma\Sigma^\ast$ and $\hat{u}$ its associated eigenvector. 

Obviously, since $\omega_k$ is not known, neither is $\rho_k$. Therefore, we cannot proceed here to localization based on the fluctuations of $\hat{\lambda}$. Instead, we will use $\hat{\lambda}$ precisely as an estimate of $\rho_k$, which we know is consistent with growing $N,n$. From $\hat{\lambda}$, assumed larger than $(1+\sqrt{c})^2$, we want to derive an estimate $\hat{\omega}$ of $\omega_k$ ($k$ is the effective failure index). This is obtained from an inversion of the relation \eqref{eq:rhoMP}. Precisely, we obtain
\begin{equation*}
	\hat{\omega} \triangleq \frac12(\hat{\lambda} - (1+c)) + \frac12 \sqrt{(\hat{\lambda}-(1+c))^2-4c}
\end{equation*}
if $\hat{\lambda}>(1+\sqrt{c})^2$ and
\begin{equation*}
	\hat{\omega} \triangleq \frac12(\hat{\lambda} - (1+c)) - \frac12 \sqrt{(\hat{\lambda}-(1+c))^2-4c}
\end{equation*}
if $\hat{\lambda}<(1-\sqrt{c})^2$.

From this estimate, we then obtain an estimate $\hat{\zeta}$ of $\zeta_k$ as follows
\begin{equation*}
	\hat{\zeta} = \frac{1-c\hat{\omega}^{-2}}{1+c\hat{\omega}^{-1}}.
\end{equation*}

A natural object to consider for the failure localization is now $|u_k^\ast \hat{u}|^2-\hat{\zeta}$. To provide a diagnosis test, we need to derive the fluctuations of this random variable. From Theorem \ref{th:2ord}, the fluctuations of $\sqrt{N}(|u_k^\ast \hat{u}|^2 - \zeta_k)$ depend on $\omega_k$ but not on $u_k$. From the expression of $\hat{\zeta}$, it is immediate that the fluctuations of $\sqrt{N}(\hat{\zeta} - \zeta_k)$ also depend on $\omega_k$ only. But since $\omega_k$ is estimated by $\hat\omega$, irrespective of the failure index $k$, the diagnosis test leads to finding the most likely argument $\hat{k}'$ among $K$ variables with same Gaussian statistics. This therefore simplifies the estimator $\hat{k}'$ of the most likely index $k$ to the following minimum-distance estimator
\begin{equation*}
	\hat{k}' = \arg\min_{k\in \{1,\ldots,K\}} \left||u_k^\ast \hat{u}|^2-\hat{\zeta}\right|.
\end{equation*}

Note importantly that, contrary to our proposed scheme, the optimal maximum-likelihood localization method cannot be easily extended to the scenario of unknown failure amplitude, therefore bringing another significant advantage of the subspace approach.

In the next section, we provide simulation results for single failure localization for the detection and localization algorithms assuming the failure amplitude known or unknown, applied to the scenarios of Section \ref{sec:ex1} and Section \ref{sec:ex2}, respectively.

 \subsection{Simulations}

\begin{figure}
	\centering
	\begin{tikzpicture}[scale=0.75,transform shape]
		\Vertex[x=0,y=0]{1}
		\node[blue] at (0,-0.7) {$({\bf 2.36})$};
		\Vertex[x=1,y=2]{2}
		\node[blue] at (1,2.7) {$({\bf 3.31})$};
		\Vertex[x=2,y=0]{3}
		\node[blue] at (2,-0.7) {$({\bf 4.50})$};
		\Vertex[x=3,y=2]{4}
		\node[blue] at (3,2.7) {$({\bf 4.25})$};
		\Vertex[x=4,y=0]{5}
		\node[blue] at (4,-0.7) {$({\bf 4.12})$};
		\Vertex[x=5,y=2]{6}
		\node[blue] at (5,2.7) {$({\bf 4.29})$};
		\Vertex[x=6,y=0]{7}
		\node[blue] at (6,-0.7) {$({\bf 4.43})$};
		\Vertex[x=7,y=2]{8}
		\node[blue] at (7,2.7) {$({\bf 4.41})$};
		\Vertex[x=8,y=0]{9}
		\node[blue] at (8,-0.7) {$({\bf 3.71})$};
		\Vertex[x=9,y=2]{10}
		\node[blue] at (9,2.7) {$({\bf 2.82})$};
		\tikzstyle{LabelStyle}=[fill=white]
		\Edge[label=$0.78$](1)(2)
		\Edge[label=$0.91$](1)(3)
		\Edge[label=$0.92$](2)(3)
		\Edge[label=$0.80$](2)(4)
		\Edge[label=$0.81$](3)(4)
		\Edge[label=$0.83$](3)(5)
		\Edge[label=$0.78$](4)(5)
		\Edge[label=$0.77$](4)(6)
		\Edge[label=$0.85$](5)(6)
		\Edge[label=$0.95$](5)(7)
		\Edge[label=$0.96$](6)(7)
		\Edge[label=$0.84$](6)(8)
		\Edge[label=$0.82$](7)(8)
		\Edge[label=$0.96$](7)(9)
		\Edge[label=$0.99$](8)(9)
		\Edge[label=$0.96$](8)(10)
		\Edge[label=$0.99$](9)(10)
	\end{tikzpicture}
	\caption{Network of $N=10$ sensors. The correlation $\EE[y(i)^\ast y(j)]$ between data on sensors $i$ and $j$, $i\neq j$, can be read on the link $(i,j)$, while $\EE[|y(i)|^2]$ variances are shown in parentheses.}
	\label{fig:sensor_network}
\end{figure}
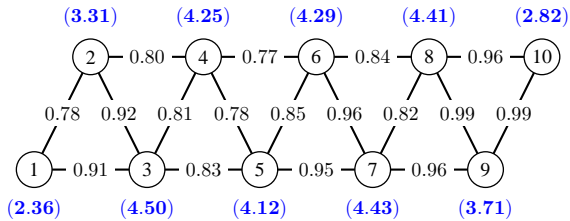

In this section, we focus on the application of the algorithms designed in Sections \ref{sec:detect_algo} and \ref{sec:localize_algo} for single node failure in the scenario of Section \ref{sec:ex1} and single parameter change in the scenario of Section \ref{sec:ex2}. 

 \subsubsection{Node failure in the scenario of Section \ref{sec:ex1}}
 Our first application example relates to the sensor network model $y=H\theta+\sigma w$ of Section \ref{sec:ex1} for $N=10$ nodes, $p=N$, and $\sigma^2=-20~{\rm dB}$. This is depicted in Figure \ref{fig:sensor_network}, where the entries of $HH^\ast+\sigma^2I_N$ are presented. We also take $\sigma_k^2=\sum_{i=1}^N(HH^\ast)_{ki}$, which is a natural assumption to avoid that a mere energy detector on $y(k)$ provide a simpler solution to our problem. This failure amplitude is assumed known by the experimenter. In practical scenarios, this may arise if a sensor starts returning time delayed data, supposedly uncorrelated with real-time data but with same variance. We assume a single failure scenario. In this context, it appears that, for all $k$, $\omega_{k,1}>0$, $\omega_{k,2}<0$ and $\omega_{k,1}$ is much larger than $|\omega_{k,2}|$. It is therefore more interesting only to consider the largest eigenvalue of $\Sigma\Sigma^\ast$ to detect and locate an hypothetical node failure. Under these conditions, the theoretical threshold for $c_N=N/n$ (if $N,n$ were large) is $0.8$ with the worst-case failure corresponding to a failure of node $10$. We therefore carry out $100\,000$ Monte Carlo simulations of node $10$ failures for $n$ varying from $8$ to $140$ and under false alarm rates varying from $10^{-2}$ to $10^{-4}$. This is depicted in Figure \ref{fig:simu10}, where it can be observed that, for $n=8$, detection and localization are barely possible, although it is clearly the starting point where detection becomes feasible. For not too large $n$, while detection rates increase, we observe that localization capabilities are still unsatisfying. This is mainly due to the inappropriate fit of the large dimensional model with $N=10$ and with the eigenvectors corresponding to the extreme eigenvalues of $\Sigma\Sigma^\ast$ being too loosely correlated to their associated population eigenvectors. Larger values of $n$ show much better performance with miss localization probability going to zero as $n\to\infty$. In particular, about five times the asymptotically optimal ratio $n/N$ is required for localization to be very efficient. In this case, the large dimensional model for the fluctuations of the eigenvalues and eigenvectors is more adapted.

\begin{figure}
  \centering
  \begin{tikzpicture}[font=\footnotesize,scale=1]
    \renewcommand{\axisdefaulttryminticks}{8} 
    \tikzstyle{every major grid}+=[style=densely dashed]       
    \tikzstyle{every axis legend}+=[cells={anchor=west},fill=white,
        at={(0.98,0.02)}, anchor=south east, font=\scriptsize ]

    \begin{axis}[
      xmajorgrids=true,
      xlabel={$n$},
      ylabel={Correct detection/localization rates},
      xmin=85,
      xmax=422, 
      ymin=0, 
      ymax=1,
      ]

      \addplot[smooth,dashed,mark=x,red,line width=0.5pt] plot coordinates{
      (85,0.000300)(106,0.028600)(127,0.063900)(148,0.127700)(169,0.215000)(190,0.324600)(211,0.459000)(232,0.587000)(253,0.709700)(274,0.800700)(295,0.873900)(316,0.925700)(337,0.959500)(358,0.977800)(379,0.988800)(400,0.995400)(422,0.998100)
      };
      \addplot[smooth,blue,mark=x,line width=0.5pt] plot coordinates{
      (85,0.000000)(106,0.000000)(127,0.013200)(148,0.123300)(169,0.214900)(190,0.324400)(211,0.459000)(232,0.587000)(253,0.709700)(274,0.800700)(295,0.873900)(316,0.925700)(337,0.959500)(358,0.977800)(379,0.988800)(400,0.995400)(422,0.998100)
      };
      \addplot[smooth,dashed,mark=*,red,line width=0.5pt] plot coordinates{
      (85,0.000500)(106,0.075300)(127,0.136900)(148,0.229700)(169,0.348100)(190,0.487100)(211,0.606600)(232,0.726900)(253,0.822600)(274,0.892700)(295,0.937000)(316,0.966700)(337,0.983600)(358,0.991100)(379,0.996900)(400,0.998600)(422,0.999800)
      };
      \addplot[smooth,mark=*,blue,mark=*,line width=0.5pt] plot coordinates{
      (85,0.000000)(106,0.000000)(127,0.013200)(148,0.184400)(169,0.345800)(190,0.487000)(211,0.606600)(232,0.726900)(253,0.822500)(274,0.892700)(295,0.937000)(316,0.966700)(337,0.983600)(358,0.991100)(379,0.996900)(400,0.998600)(422,0.999800)
      };
      \addplot[smooth,dashed,mark=triangle,red,line width=0.5pt] plot coordinates{
      (85,0.000600)(106,0.189800)(127,0.289500)(148,0.411400)(169,0.549400)(190,0.667300)(211,0.774000)(232,0.864500)(253,0.918700)(274,0.957100)(295,0.975900)(316,0.988200)(337,0.992900)(358,0.998100)(379,0.999200)(400,0.999400)(422,0.999800)
      };
      \addplot[smooth,mark=*,blue,mark=triangle,line width=0.5pt] plot coordinates{
      (85,0.000000)(106,0.000000)(127,0.017000)(148,0.207900)(169,0.525600)(190,0.665300)(211,0.773800)(232,0.864300)(253,0.918700)(274,0.957100)(295,0.975900)(316,0.988200)(337,0.992900)(358,0.998100)(379,0.999200)(400,0.999400)(422,0.999800)
      };
      \legend{{CDR, FAR$=10^{-4}$},{CLR, FAR$=10^{-4}$},{CDR, FAR$=10^{-3}$},{CLR, FAR$=10^{-3}$},{CDR, FAR$=10^{-2}$},{CLR, FAR$=10^{-2}$}}
    \end{axis}
  \end{tikzpicture}
  \caption{Correct detection (CDR) and localization (CLR) rates for different levels of false alarm rates (FAR) and different values of $n$, for worst case node failure in a $100$-node sensor network. The minimal theoretical $n$ for observability is $n=85$.}
  \label{fig:simu100}
\end{figure}
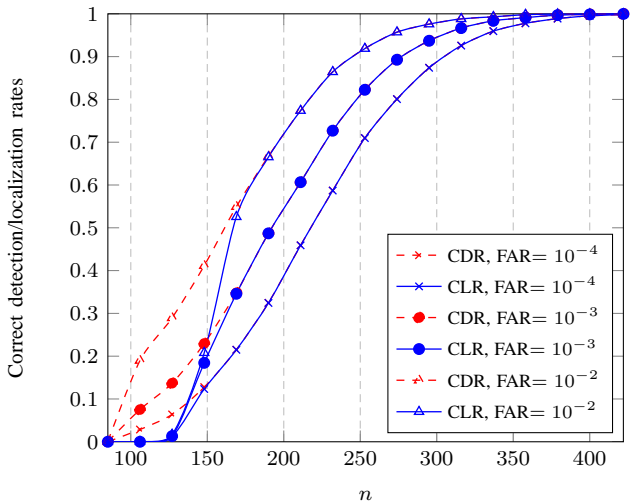

The same conditions are simulated for a system with $N=100$ nodes in which each node has eight neighbors and with correlation values of the same order of magnitude as in Figure \ref{fig:sensor_network}. The detectability threshold for $N/n$ is here $0.85$ and we still consider the worst case failure scenario. This is depicted in Figure \ref{fig:simu100}, where one can see that smaller ratios $n/N$ over the asymptotically optimal threshold are demanded for high detectability and localization ability to appear, when compared to the scenario $N=10$.

\begin{figure}
  \centering
  \begin{tikzpicture}[font=\footnotesize,scale=1]
    \renewcommand{\axisdefaulttryminticks}{8} 
    \tikzstyle{every major grid}+=[style=densely dashed]       
    \tikzstyle{every axis legend}+=[cells={anchor=west},fill=white,
        at={(0.98,0.02)}, anchor=south east, font=\scriptsize ]

    \begin{axis}[
      xmajorgrids=true,
      xlabel={$n$},
      ylabel={Correct detection/localization rates},
      xmin=8,
      xmax=140, 
      ymin=0, 
      ymax=1,
      ]

      \addplot[smooth,dashed,mark=x,red,line width=0.5pt] plot coordinates{
      (8,0.010200)(16,0.043800)(24,0.107960)(32,0.207860)(40,0.336030)(47,0.454400)(55,0.584990)(63,0.700300)(71,0.794630)(79,0.862870)(86,0.909710)(94,0.944260)(102,0.966340)(110,0.981400)(118,0.989790)(125,0.994050)(133,0.996800)(141,0.998470)(149,0.999110)(157,0.999680)(164,0.999820)
      };
      \addplot[smooth,blue,mark=x,line width=0.5pt] plot coordinates{
      (8,0.000000)(16,0.010520)(24,0.096780)(32,0.202810)(40,0.333010)(47,0.452000)(55,0.582830)(63,0.698180)(71,0.792920)(79,0.861600)(86,0.908710)(94,0.943300)(102,0.965540)(110,0.980880)(118,0.989350)(125,0.993700)(133,0.996450)(141,0.998210)(149,0.998970)(157,0.999550)(164,0.999620)
      };
      \addplot[smooth,dashed,mark=*,red,line width=0.5pt] plot coordinates{
      (8,0.026660)(16,0.093370)(24,0.202310)(32,0.338200)(40,0.487050)(47,0.608330)(55,0.728300)(63,0.820520)(71,0.885220)(79,0.930940)(86,0.958050)(94,0.976350)(102,0.987440)(110,0.993380)(118,0.996930)(125,0.998020)(133,0.999030)(141,0.999610)(149,0.999840)(157,0.999940)(164,0.999940)
      };
      \addplot[smooth,mark=*,blue,mark=*,line width=0.5pt] plot coordinates{
      (8,0.000000)(16,0.024870)(24,0.173410)(32,0.326340)(40,0.481210)(47,0.603840)(55,0.724740)(63,0.817780)(71,0.883170)(79,0.929270)(86,0.956680)(94,0.975450)(102,0.986740)(110,0.992800)(118,0.996450)(125,0.997750)(133,0.998750)(141,0.999310)(149,0.999710)(157,0.999800)(164,0.999730)
      };
      \addplot[smooth,dashed,mark=triangle,red,line width=0.5pt] plot coordinates{
      (8,0.058580)(16,0.209900)(24,0.366650)(32,0.529530)(40,0.678360)(47,0.776730)(55,0.860030)(63,0.919240)(71,0.954200)(79,0.975700)(86,0.986000)(94,0.992830)(102,0.996510)(110,0.998370)(118,0.999240)(125,0.999570)(133,0.999880)(141,0.999920)(149,0.999980)(157,0.999970)(164,1.000000)
      };
      \addplot[smooth,mark=*,blue,mark=triangle,line width=0.5pt] plot coordinates{
      (8,0.000000)(16,0.039970)(24,0.285590)(32,0.499420)(40,0.664690)(47,0.768560)(55,0.854330)(63,0.915330)(71,0.951510)(79,0.973860)(86,0.984460)(94,0.991870)(102,0.995710)(110,0.997770)(118,0.998890)(125,0.999170)(133,0.999620)(141,0.999650)(149,0.999820)(157,0.999830)(164,0.999870)
      };
      \legend{{CDR, FAR$=10^{-4}$},{CLR, FAR$=10^{-4}$},{CDR, FAR$=10^{-3}$},{CLR, FAR$=10^{-3}$},{CDR, FAR$=10^{-2}$},{CLR, FAR$=10^{-2}$}}
    \end{axis}
  \end{tikzpicture}
  \caption{Correct detection (CDR) and localization (CLR) rates for different levels of false alarm rates (FAR) and different values of $n$, for node $10$ failure in the sensor network of Figure \ref{fig:sensor_network}. The minimal theoretical $n$ for observability is $n=8$.}
  \label{fig:simu10}
\end{figure}
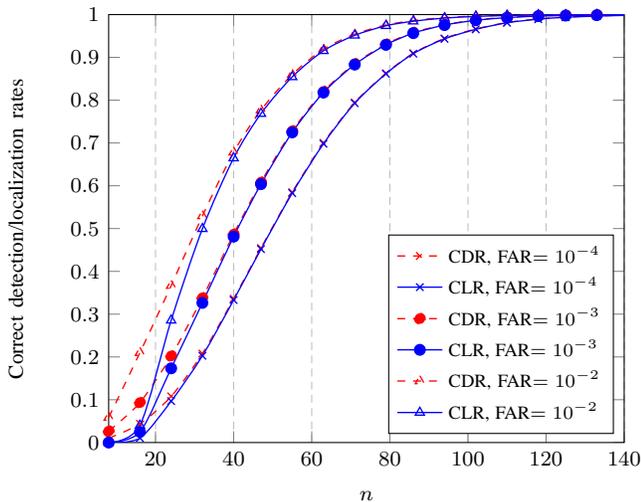

\subsubsection{Sudden unknown parameter change}

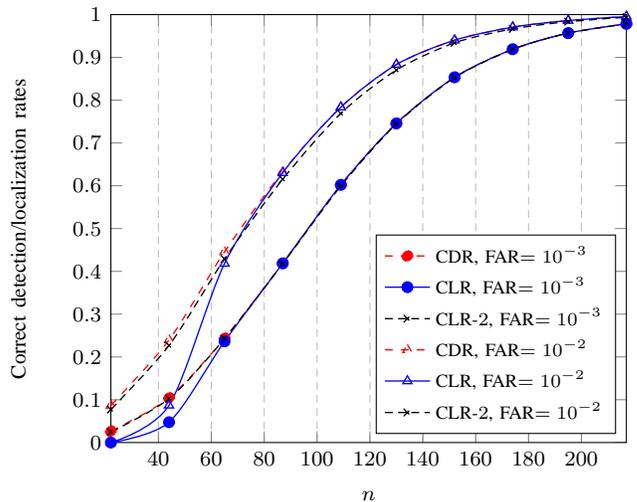
\begin{figure}
  \centering
  \begin{tikzpicture}[font=\footnotesize,scale=1]
    \renewcommand{\axisdefaulttryminticks}{8} 
    \tikzstyle{every major grid}+=[style=densely dashed]       
    \tikzstyle{every axis legend}+=[cells={anchor=west},fill=white,
        at={(0.98,0.02)}, anchor=south east, font=\scriptsize ]

    \begin{axis}[
      xmajorgrids=true,
      xlabel={$n$},
      ylabel={Correct detection/localization rates},
      xmin=22,
      xmax=217, 
      ymin=0, 
      ymax=1,
      ]

      \addplot[smooth,dashed,mark=*,red,line width=0.5pt] plot coordinates{
      (22,0.026600)(44,0.104500)(65,0.244000)(87,0.419300)(109,0.602400)(130,0.745700)(152,0.853600)(174,0.919100)(195,0.956800)(217,0.978500)
      };
      \addplot[smooth,mark=*,blue,mark=*,line width=0.5pt] plot coordinates{
      (22,0.000000)(44,0.047500)(65,0.236400)(87,0.418600)(109,0.602100)(130,0.745700)(152,0.853500)(174,0.919100)(195,0.956800)(217,0.978500)
      };
      \addplot[smooth,densely dashed,black,mark=x,line width=0.5pt] plot coordinates{
      (22,0.024600)(44,0.102600)(65,0.242800)(87,0.417900)(109,0.599400)(130,0.744400)(152,0.852500)(174,0.918600)(195,0.956700)(217,0.978200)
      };
      \addplot[smooth,dashed,mark=triangle,red,line width=0.5pt] plot coordinates{
      (22,0.086700)(44,0.239800)(65,0.445600)(87,0.632700)(109,0.783700)(130,0.883300)(152,0.940200)(174,0.970600)(195,0.986100)(217,0.995400)
      };
      \addplot[smooth,mark=*,blue,mark=triangle,line width=0.5pt] plot coordinates{
      (22,0.000000)(44,0.087000)(65,0.418400)(87,0.630300)(109,0.783600)(130,0.883200)(152,0.940100)(174,0.970600)(195,0.986100)(217,0.995400)
      };
      \addplot[smooth,densely dashed,black,mark=x,line width=0.5pt] plot coordinates{
      (22,0.077200)(44,0.226900)(65,0.428900)(87,0.615300)(109,0.769100)(130,0.870400)(152,0.934000)(174,0.966500)(195,0.983500)(217,0.993700)
      };
      \legend{{CDR, FAR$=10^{-3}$},{CLR, FAR$=10^{-3}$},{CLR-$2$, FAR$=10^{-3}$},{CDR, FAR$=10^{-2}$},{CLR, FAR$=10^{-2}$},{CLR-$2$, FAR$=10^{-2}$}}
    \end{axis}
  \end{tikzpicture}
  \caption{Correct detection (CDR) and localization rates for different levels of false alarm rates (FAR) and different values of $n$, for sudden change of parameter $10$ in the Scenario of Section \ref{sec:ex2}. Comparison is made between localization assuming $\beta_k$ known (CLR) and localization assuming $\beta_k$ unknown (CLR-$2$). The minimal theoretical $n$ for observability is $n=22$.}
  \label{fig:simu10_link}
\end{figure}

In this section, we consider the parameter change scenario of Section \ref{sec:ex2}. We still consider the network of Figure \ref{fig:sensor_network}, and $\sigma^2=-20~{\rm dB}$ as above. We now assume a sudden change of parameter $\theta(10)$ with $\beta_{10}=2$, being the worst case scenario for failure identification if $\beta_k=2$ for all $k$. We depict the performance of the failure detection and localization algorithms and compare the settings where $\beta_k$ is known or unknown in advance to the experimenter. In the former scenario, we apply the localization algorithm of Section \ref{sec:localize_algo} based on the joint fluctuations of the extreme eigenvalues and eigenspace projections, while in the latter, we apply the localization algorithm of Section \ref{sec:unknown_failure}, where a prior step of eigenvalue inference is performed before the study of the fluctuations of the eigenspace projections. The results are presented in Figure \ref{fig:simu10_link}.

It appears from Figure \ref{fig:simu10_link} that the suboptimal algorithm of Section \ref{sec:unknown_failure} performs only slightly worse than the algorithm of Section \ref{sec:localize_algo} for large $n$, and that it even performs better for small $n$. This last observation is explained by the inadequacy of the theoretical value of $\zeta_k$ for too small values of $n$. It is therefore interesting to see that, for practical purposes, the absence of prior knowledge on the amplitude of the failure does not severely reduce the efficiency of the localization algorithm.

 \section{Conclusion}
 \label{sec:conclusion}
In this article, a characterization of the joint fluctuations of the extreme eigenvalues and corresponding eigenspace projections of a certain class of random matrices is provided. This characterization was used to perform fast and computationally reasonable detection and localization of multiple failures in large sensor networks through a general hypothesis testing framework. The main practical outcomes of this article lie first in a characterization of the minimum number of observations necessary to ensure failure detectability in large networks and second in the design of flexible but simple algorithms that can be adapted to multiple types of failure scenarios consistent with the small rank perturbation random matrix model. We also extend the detection and diagnosis approach to scenarios where the amplitudes of the hypothetical failures are not a priori known. Practical simulations suggest that the proposed algorithms allow for high failure detection and localization performance even for networks of small sizes, although for those much more observations than theoretically predicted are in general demanded. 

\appendices

\section{Proofs of results of Section \ref{sec:main_res}} 

\subsection{Proof of Lemma \ref{lm:cltvep}}
\label{anx:cltvep} 
We shall assume without loss of generality that $i=1$. In Section \ref{sec:vep}, we saw that 
\begin{equation*}
U_1^* \widehat\Pi_1 U_1 
= \frac{1}{2\pi\imath} \oint_{\gamma_1} 
\widehat A_1(z)^* \widehat H(z)^{-1} \widehat A_2(z) dz  
\end{equation*}
with probability one, where 
\begin{align*} 
\widehat{A}_1(z)^\ast &= z U_1^\ast(I_N+P)^{-\oh} Q(z)U  \\
&= \frac{z}{(1+\omega_1)^{\frac12}} U_1^\ast Q(z)U
\end{align*}
and
\begin{align*} 
\widehat{A}_2(z) &= \Omega(I_r+\Omega)^{-1} U^\ast Q(z)(I_N+P)^{-\oh} U_1 \\
&= (1+\omega_1)^{-\frac12} \Omega(I_r+\Omega)^{-1}U^\ast Q(z) U_1 
\end{align*} 
(take $b_1$ and $b_2$ as any two columns of $U_1$ in \eqref{bpib}). 
Similarly, 
\begin{equation*}
\zeta_1 I_{j_1} = \frac{1}{2\pi\imath} \oint_{\gamma_1} 
A_1(z)^* H(z)^{-1} A_2(z) dz  
\end{equation*}
where 
\begin{align*} 
{A}_1(z)^\ast &= zm(z) U_1^* (I+P)^{-\frac12} U \\ 
&= \frac{z m(z)}{(1+\omega_1)^{\frac12}} 
\begin{bmatrix} I_{j_1} & 0 \end{bmatrix} , \\
{A}_2(z) &= 
m(z) \Omega(I+\Omega)^{-1} U^* (I+P)^{-\frac12} U_1 \\ 
&= \frac{\omega_1 m(z)}{(1+\omega_1)^{3/2}}  
\begin{bmatrix} I_{j_1} \\ 0 \end{bmatrix} . 
\end{align*} 
Fixing $z \in \gamma_1$, we have 
\begin{equation*}
\widehat H(z)^{-1} = H(z)^{-1} - H(z)^{-1} E(z) H(z)^{-1} + {\mathcal O}(\| E(z) \|^2)
\end{equation*}
where 
\begin{equation*}
E(z) = \widehat H(z) - H(z) = z \Omega(I_r+\Omega)^{-1} U^* (Q(z) - m(z) I) U.
\end{equation*}
Guided by this observation, we write 
\begin{align*}
V_{1,n} &=  
\frac{\sqrt{N}}{2\pi\imath} \oint_{\gamma_1} 
\left( 
\widehat A_1(z)^* \widehat H(z)^{-1} \widehat A_2(z) \right. \\ 
& 
\ \ \ \ 
\left. \phantom{\widehat H(z)^{-1}} 
- A_1(z)^* H(z)^{-1} A_2(z) 
\right) \, dz  \quad (\text{a.s., } n \ \text{large}) \\
&=   
\frac{\sqrt{N}}{2\pi\imath} \oint_{\gamma_1} 
(\widehat A_1(z)^* - A_1(z)^* ) H(z)^{-1} A_2(z) \, dz  \\
&\phantom{=} 
+ \frac{\sqrt{N}}{2\pi\imath} \oint_{\gamma_1} 
A_1(z)^* H(z)^{-1} (\widehat A_2(z) - A_2(z)) \, dz  \\
&\phantom{=} 
- \frac{\sqrt{N}}{2\pi\imath} \oint_{\gamma_1} 
A_1(z)^* H(z)^{-1}E(z) H(z)^{-1} A_2(z) \, dz  \\
&\phantom{=} + \mathcal{E} \\
&= Z_1 + Z_2 + Z_3 + \mathcal{E} 
\end{align*} 
where $\mathcal E$ contains all the higher order terms that appear when we develop the integrand at the right hand side of the first equality.

In what follows, we successively study each of the terms at the right hand side of this equation. Recalling \eqref{eq:H-1}, the term $Z_1$ writes 
\begin{equation*}
Z_1 = \frac{\sqrt{N}}{2\pi\imath} \oint_{\gamma_1} 
\frac{z m(z) \, U_1^* (Q(z) - m(z) I) U_1}{(1+\omega_1) 
\left( (1+\omega_1) / \omega_1 + zm(z) \right) } dz . 
\end{equation*}
The denominator has one simple zero in $\interior(\gamma_1)$. With probability one, the numerator has no zero in $\interior(\gamma_1)$. Using the residue theorem and the identity $(1+\omega_1)^{-1} = (1+h(\rho_1))/h(\rho_1)$, we obtain 
\begin{equation*}
Z_1 = \frac{1 + h(\rho_1)}{h'(\rho_1)} \, \sqrt{N} 
U_1^* (Q(\rho_1) - m(\rho_1) I) U_1 . 
\end{equation*}
Similarly, the term $Z_2$ writes 
\begin{equation*}
Z_2 =  Z_1 = \frac{1 + h(\rho_1)}{h'(\rho_1)} \, \sqrt{N} 
U_1^* (Q(\rho_1) - m(\rho_1) I) U_1 . 
\end{equation*}
Turning to $Z_3$, we have 
\begin{equation*}
Z_3 = 
- \frac{\sqrt{N}}{2\pi\imath} \oint_{\gamma_1}  
\frac{z^2 m(z)^2 \ U_1^* (Q(z) - m(z)) U_1 } 
{ (1+\omega_1)\left((1+\omega_1)/\omega_1 + z m(z)\right)^2 } dz 
\end{equation*}
which shows that we have a pole with degree $2$ in $\interior(\gamma_1)$. Write the integrand as $G(z)/g(z)$ and recall that the residue of a meromorphic function $f(z)$ associated with a degree $2$ pole at $z_0$ is $\lim_{z\to z_0} d\left( (z-z_0)^2 f(z) \right) / dz$.
After some simple calculations, this results in 
\begin{align*}
Z_3 
&=  
\frac{G(\rho_1) g''(\rho_1)}{g'(\rho_1)^3} 
- \frac{G'(\rho_1)}{g'(\rho_1)^2}  \\
&= 
\frac{1+h(\rho_1)}{h'(\rho_1)} 
\left( 
\frac{ h(\rho_1) h''(\rho_1)}
{h'(\rho_1)^2} - 2 
\right) 
\\ 
& \times\sqrt{N} U_1^* (Q(\rho_1) - m(\rho_1)I) U_1 \\ 
&- 
\frac{h(\rho_1)(1+h(\rho_1))}{ h'(\rho_1)^2}  \, 
\sqrt{N} U_1^* (Q'(\rho_1) - m'(\rho_1)I) U_1 .  
\end{align*} 

We now show briefly that the last term $\mathcal E$ in the expression of $V_{1,n}$ converges to zero in probability. A more detailed argument is given in \cite{hac-spike-11}. Recall that $\mathcal E$ accounts for all the higher order terms that show up when we expand the integrand $\widehat A_1 \widehat H^{-1} \widehat A_2 - A_1 H^{-1} A_2$. Let us focus on one of these terms, namely 
\begin{align*} 
&\frac{\sqrt{N}}{2\pi\imath} \oint_{\gamma_1} A_1(z)^* \\
&\times \left( \widehat H(z)^{-1} - H(z)^{-1} + H(z)^{-1} E(z) H(z)^{-1} \right) A_2(z)dz 
\end{align*} 
and show that it converges in probability to zero. The other terms can be treated similarly. First, we can show that 
\begin{equation*}
\| \widehat H(z)^{-1} - H(z)^{-1} + H(z)^{-1} E(z) H(z)^{-1} \| \leq K \| E(z) \|^2
\end{equation*}
on $\gamma_1$ where $K$ is some constant. Now we write 
\begin{align*}
	E(z) &= z \Omega(I_r+\Omega)^{-1} U^* (Q(z) - \alpha(z) I) U \\
	&+ z (\alpha(z) - m(z)) \Omega(I_r+\Omega)^{-1} \\
	&= E_1(z) + E_2(z).
\end{align*}
	Noticing that $A_1(z)$ and $A_2(z)$ are bounded on $\gamma_1$, and writing $z = \rho_1 + R \exp(2\imath\pi\theta)$ on $\gamma_1$, the result is shown if we show that 
\begin{equation}
\label{cvg-prob} 
\sqrt{N} \int_0^1 \| E_i(\rho_1+Re^{2\imath\pi\theta}) \|^2 d\theta \probto 0 
\end{equation} 
for $i=1,2$. Lemma \ref{lm-fq} shows that $\EE  \| E_1(z) \|^2 \leq K' / n$ on $\gamma_1$ where the constant $K'$ is independent of $z$. By Markov's inequality, \eqref{cvg-prob} is true for $E_1$. Convergence for $E_2$ is obtained from Assumption {\bf A4} in conjunction with the analyticity of $\alpha(z) - m(z)$, as shown in \cite{hac-spike-11}.

Taking the sum $Z_1 + Z_2 + Z_3$, we obtain the desired result. 

\subsection{Proof of Lemma \ref{lm:vap}} 
\label{anx:vap} 
Set $i=1$ and write 
\begin{equation*}
\Omega(I_r+\Omega)^{-1} = 
\begin{bmatrix} 
\beta_1 I_{j_1} \\ & B_1 
\end{bmatrix} . 
\end{equation*}
Let $y_n = \rho_1 + x/\sqrt{N}$. Write $U = \begin{bmatrix} U_1 & \widetilde U_1 \end{bmatrix}$ and 
\begin{align*} 
\widehat H(y_n) &= 
\begin{bmatrix}
I_{j_1} + \beta_1 y_n U_1^* Q(y_n) U_1 & 
\!\!\! 
\beta_1 y_n U_1^* Q(y_n) \widetilde U_1 \\
y_n B_1 \widetilde U_1^* Q(y_n) U_1 & 
\!\!\!\!\!
I_{r-j_1} + y_n B_1 \widetilde U_1^* Q(y_n) \widetilde U_1  
\end{bmatrix} \\
&= 
\begin{bmatrix} \widehat H_{11} & \widehat H_{12} \\
\widehat H_{21} & \widehat H_{22} 
\end{bmatrix} .  
\end{align*} 

Let us study 
\begin{align*}
	N^{\frac{j_1}2} \det \widehat H(y_n) = \det(\widehat H_{22}) \det( \sqrt{N} \widehat H_{11} - \sqrt{N} \widehat H_{12}\widehat H_{22}^{-1} \widehat H_{21}).
\end{align*}
	By Lemma \ref{lm-fq}, 
\begin{equation*}
\widehat H_{22} \asto 
\begin{bmatrix}
(1-\beta_2/\beta_1) I_{j_2} & \\ 
& \ddots \\
& & (1-\beta_t/\beta_1) I_{j_t} 
\end{bmatrix} 
\end{equation*}
hence
\begin{equation*}
\det\widehat H_{22} \asto \prod_{\ell>1} (1 - \beta_\ell / \beta_1)^{j_\ell}.
\end{equation*}
By the same lemma, 
\begin{equation*}
\EE\left[ \sqrt{N} \1_{\sigma(XX^\ast) \in [a-\varepsilon, b+\varepsilon]} \| \widehat H_{12} \|^2\right] \leq K/\sqrt{N}
\end{equation*}
and 
\begin{equation*}
	\EE \left[\1_{\sigma(XX^\ast) \in [a-\varepsilon, b+\varepsilon]} \| \widehat H_{21} \|\right] \leq K'/\sqrt{N}
\end{equation*}
where $K$ and $K'$ are some constants, hence 
\begin{equation*}
\det( \sqrt{N} \widehat H_{11} - \sqrt{N} \widehat H_{12}\widehat H_{22}^{-1} \widehat H_{21}) -  \det( \sqrt{N} \widehat H_{11}) \probto 0.
\end{equation*}
Let us study this last term. We write
\begin{align*} 
\sqrt{N} \widehat H_{11} &= 
\sqrt{N} \beta_1 y_n U_1^* ( Q(y_n)  - \alpha(y_n) I_N) U_1 \\
&\phantom{=} 
+ \sqrt{N} \beta_1 (y_n \alpha(y_n) - \rho_1 \alpha(\rho_1)) I_{j_1} \\
&\phantom{=} 
+ \sqrt{N} \beta_1 (\rho_1 \alpha(\rho_1) - \rho_1 m(\rho_1)) I_{j_1} \\ 
&= Z_1 + Z_2 + Z_3 . 
\end{align*} 
Writing
\begin{align*}
	Q(y_n) - Q(\rho_1) &= Q(y_n) (Q(\rho_1)^{-1} - Q(y_n)^{-1}) Q(\rho_1) \\
	&= N^{-\frac12} x Q(y_n) Q(\rho_1)
\end{align*}
we have 
\begin{align*} 
&Z_1 - \sqrt{N} \beta_1 y_n U_1^* ( Q(\rho_1)  - \alpha(\rho_1) I) U_1 \\ 
&= x \beta_1 y_n U_1^* (  Q(y_n) Q(\rho_1) - N^{-1} (\tr Q(y_n) Q(\rho_1)) I ) U_1 \\ 
&\probto 0 
\end{align*} 
by Lemma \ref{lm-fq}. From {\bf A4}, we further have 
\begin{equation*}
Z_1 - \sqrt{N} \beta_1 \rho_1 U_1^* ( Q(\rho_1)  - m(\rho_1) I) U_1 
\probto 0 . 
\end{equation*}
Turning to the second term, we can show using {\bf A4} that 
\begin{equation*}
Z_2 = \beta_1 x 
\frac{ y_n \alpha(y_n) -  \rho_1 \alpha(\rho_1) }
{x/\sqrt{N}} 
\asto \beta_1 x (\rho_1 m(\rho_1))' .
\end{equation*}
Again by {\bf A4}, $Z_3 \asto 0$. This results in 
\begin{align*} 
&N^{j_1/2} \det \widehat H(\rho_1 + x/\sqrt{N}) - \Bigl( \prod_{\ell>1} (1 - \beta_\ell / \beta_1)^{j_\ell} \Bigr)\\ 
&\times \det\left( 
\sqrt{N} \beta_1 \rho_1 U_1^* ( Q(\rho_1)  - m(\rho_1) I) U_1 
+ 
\beta_1 x h'(\rho_1) \right) \\ 
&\probto 0 . 
\end{align*}
The same argument for $i > 1$ leads to the result.  

\subsection{Proof of Lemma \ref{lm:tlc}}
\label{anx:tlc} 

Recall that $XX^*$ admits the spectral factorization $XX^* = W \Lambda W^*$ where $W$ and $\Lambda = \diag(\lambda_1, \ldots, \lambda_N)$ are independent, and where $W$ is Haar distributed on the group of $N\times N$ unitary matrices. 

From Assumption {\bf A4} and the analyticity of $f_1$, we can show as in \cite{BAI04} that 
\begin{equation*}
\sqrt{N} \left( 
\frac 1N \sum_{k=1}^N f_1(\lambda_k) - \int f_1(\lambda) d\pi(\lambda) \right) 
\probto 0  
\end{equation*}
hence the lemma is shown if we show the result on 
\begin{equation*}
\underline{S}_n \! = \! \left( \! \sqrt{N} \begin{bmatrix} 
U_i^* f_i(XX^*) U_i - 
\left( \frac1N \sum_{k=1}^N f_i(\lambda_k) \right) I_{j_i} \\
U_i^* g_i(XX^*) U_i - \left( \frac1N \sum_{k=1}^N g_i(\lambda_k) \right) I_{j_i} 
\end{bmatrix} \right)_{i=1}^t .  
\end{equation*}
Let $Z$ be a $N \times r$ random matrix with independent ${\cal CN}(0,1)$ elements, chosen independently of $\Lambda$. Write $Z = [ Z_1 \ldots Z_t ]$ where the blocks $Z_\ell$ have the same dimensions as the $U_\ell$. Then $\underline{S}_n$ is equal in distribution to 
\begin{align*} 
\sqrt{N}
\begin{pmatrix} \begin{bmatrix} 
[Z (Z^* Z)^{-\frac12}]_i^* \left( f_i(\Lambda) - 
\frac{\tr f_i(\Lambda)I_{j_i}}{N}  \right) [Z (Z^* Z)^{-\frac12}]_i \\ 
[Z (Z^* Z)^{-\frac12}]_i^* \left( g_i(\Lambda) - 
\frac{\tr g_i(\Lambda)I_{j_i}}{N} 
\right) [Z (Z^* Z)^{-\frac12}]_i 
\end{bmatrix} \end{pmatrix}_{i=1}^t
\end{align*} 
where $[Z (Z^* Z)^{-\frac12}]_i$ is the matrix formed by the columns $j_1+\ldots+j_{i-1}$ to $j_1+\ldots+j_i$ of $Z (Z^* Z)^{-\frac12}$.
By the law of large numbers, $N^{-1} Z^* Z \asto I_r$, hence it will be enough to show the result on 
\begin{align*}
\bar{S}_n &= 
\frac{1}{\sqrt{N}} \begin{pmatrix} \begin{bmatrix} 
Z_i^* \left( f_i(\Lambda) - N^{-1}\tr f_i(\Lambda) I_N \right) Z_i \\ 
Z_i^* \left( g_i(\Lambda) - N^{-1}\tr g_i(\Lambda) I_N \right) Z_i 
\end{bmatrix} \end{pmatrix}_{i=1}^t \\ 
&= 
\frac{1}{\sqrt{N}} \sum_{k=1}^N 
\begin{pmatrix}\begin{bmatrix} 
\left( f_i(\lambda_k) - \frac{\tr f_i(\Lambda)}{N} \right)
\left( z_{i,k} z_{i,k}^* - I_{j_i} \right) \\ 
\left( g_i(\lambda_k) - \frac{\tr g_i(\Lambda)}{N} \right)
\left( z_{i,k} z_{i,k}^* - I_{j_i} \right) 
\end{bmatrix}\end{pmatrix}_{i=1}^t 
\end{align*}
where we have written $Z_i^* = [ z_{i,1} \ldots z_{i,N} ]$. Write $c_{i,k} = f_i(\lambda_k) - N^{-1}\tr f_i(\Lambda)$ and $d_{i,k} = g_i(\lambda_k) - N^{-1}\tr g_i(\Lambda)$. Up to an element rearrangement, $\bar{S}_n$ can be rewritten as the $\RR^{2(j_1^2+\cdots+j_t^2)}$-valued vector 
\begin{equation}
\label{rear} 
\frac{1}{\sqrt{N}} 
\sum_{k=1}^N 
\begin{bmatrix} 
v_{i,k} \otimes 
\begin{bmatrix} c_{i,k} \\ d_{i,k} \end{bmatrix} 
\end{bmatrix}_{i=1}^t 
\end{equation} 
where for every $i=1,\ldots,t$, the $N$ vectors $v_{i,k}$ are valued in $\RR^{j_i^2}$. We shall show that this sum converges in law to a Gaussian $\RR^{2(j_1^2+\cdots+j_t^2)}$-valued random vector whose covariance matrix is equal to the covariance matrix of \eqref{clt-gue} rearranged similarly to $\bar S_n$.

We observe that the summands of \eqref{rear} are centered and are independent conditionally to $\Lambda$. Observe also that for every $k$, the vectors $(v_{i,k})_{i=1}^t$ are independent and that the elements of each of these vectors are decorrelated. Based on {\bf A2} and {\bf A3}, we have 
\begin{align*} 
&\frac 1N \sum_{k=1}^N 
\EE \left[\left. 
\begin{bmatrix} 
v_{i,k} \otimes 
\begin{bmatrix} c_{i,k} \\ d_{i,k} \end{bmatrix} 
\end{bmatrix}_{i=1}^t 
\left(
\begin{bmatrix} 
v_{i,k} \otimes 
\begin{bmatrix} c_{i,k} \\ d_{i,k} \end{bmatrix} 
\end{bmatrix}_{i=1}^t 
\right)^\trans
\right| \Lambda \right] \\
&= 
\diag\left( 
I_{j_i} \otimes 
\left( \frac 1N \sum_{k=1}^N \begin{bmatrix} c_{i,k}^2 & c_{i,k} d_{i,k} \\
c_{i,k} d_{i,k} & d_{i,k}^2 
\end{bmatrix} \right) 
\right)_{i=1}^t  \\
&\asto 
\diag\left( 
I_{j_i} \otimes 
R_i 
\right)_{i=1}^t 
\end{align*} 
which coincides with the covariance matrix of \eqref{clt-gue} after the rearrangement.

Furthermore, thanks to {\bf A3}, it is easy to see that the Lyapunov condition 
\begin{equation*}
\frac{1}{N^{1+\eta}} 
\sum_{k=1}^N 
\EE \left[ \left. \left\|
\begin{bmatrix} 
v_{i,k} \otimes 
\begin{bmatrix} c_{i,k} \\ d_{i,k} \end{bmatrix} 
\end{bmatrix}_{i=1}^t 
\right\|^{2+2\eta} \right| 
\Lambda \right] 
\asto 0
\end{equation*}
is valid for any $\eta > 0$, hence \eqref{rear} satisfies the conditions of the central limit theorem, which proves the lemma.

\bibliography{/home/romano/phd-group/papers/rcouillet/tutorial_RMT/IEEEconf,/home/romano/phd-group/papers/rcouillet/tutorial_RMT/IEEEabrv,./tutorial_RMT,./fault_detect}

\end{document}